\documentclass[twocolumn,twoside]{IEEEtran}    

\usepackage{amsmath,amssymb}
\usepackage[utf8]{inputenc}
\usepackage{hyperref}
\usepackage[pdftex]{graphicx}
\usepackage{xspace}
\usepackage{tabularx}
\usepackage{amsfonts}
\usepackage{lmodern}
\usepackage{booktabs}

\usepackage{cite} 
\usepackage{color}

\newcommand{\LA}{\mathrm{L8}}
\newcommand{\LU}{\mathrm{LU}}
\newcommand{\PV}{\mathrm{PV}}

\begin{document}


\title{Cross-Sensor Adversarial Domain Adaptation of Landsat-8 and Proba-V images for Cloud Detection}

\author{Gonzalo Mateo-Garc{\'i}a, 
Valero Laparra, Dan L{\'o}pez-Puigdollers, Luis G{\'o}mez-Chova, {\em Senior Member, IEEE}
\thanks{Manuscript received June 2020;}
\thanks{GMG, VL, DLP and LGC are with the Image Processing Laboratory (IPL), University of Valencia, Spain. 
Web: http://isp.uv.es. E-mail: \{gonzalo.mateo-garcia,valero.laparra,dan.lopez,luis.gomez-chova\}@uv.es.}
\thanks{This work has been partially supported by the Spanish Ministry of Economy and Competitiveness (MINECO, TEC2016-77741-R, ERDF) and the European Space Agency (ESA IDEAS+ research grant, CCN008).}
}

\markboth{\scriptsize }
{\scriptsize Mateo-Garc{\'i}a et al.: Cross-Sensor Adversarial Domain Adaptation for Cloud Detection}

\maketitle
\begin{abstract}

The number of Earth observation satellites carrying optical sensors with similar characteristics is constantly growing. Despite their similarities and the potential synergies among them, derived satellite products are often developed for each sensor independently. 
Differences in retrieved radiances lead to significant drops in accuracy, which hampers knowledge and information sharing across sensors. 
This is particularly harmful for machine learning algorithms, since gathering new ground truth data to train models for each sensor is costly and requires experienced manpower. 
In this work, we propose a domain adaptation transformation to reduce the statistical differences between images of two satellite sensors in order to boost the performance of transfer learning models. 
The proposed methodology is based on the Cycle Consistent Generative Adversarial Domain Adaptation (CyCADA) framework that trains the transformation model in an unpaired manner. In particular, Landsat-8 and Proba-V satellites, which present different but compatible spatio-spectral characteristics, are used to illustrate the method. 
The obtained transformation significantly reduces differences between the image datasets while preserving the spatial and spectral information of adapted images, which is hence useful for any general purpose cross-sensor application. 
In addition, the training of the proposed adversarial domain adaptation model can be modified to improve the  performance in a specific remote sensing application, such as cloud detection, by including a dedicated term in the cost function. Results show that, when the proposed transformation is applied, cloud detection models trained in Landsat-8 data increase cloud detection accuracy in Proba-V.

\end{abstract}


\section{Introduction}\label{sec:intro}

Over the last decade, the number of both private and public satellite missions for Earth observation has explode. According to the UCS satellite database~\cite{ucsatellite}, there are currently around 500 orbiting satellites carrying passive optical sensors (multispectral or hyperspectral). Whereas each sensor is to some extent unique, in many cases there are only small differences between them such as slightly different spectral responses, different ground sampling distances, or the different inherent noise of the instruments. Nevertheless, derived products from those images are currently tailored to each particular sensor, since models designed to one sensor often transfer poorly to a different one due to those differences~\cite{DARSTuia2016}. 
In order to transfer products across sensors we need to ensure that the underlying data distribution does not change from one sensor to the other. 
In machine learning this is a long-standing problem that is called {\em data shift}~\cite{torralba_unbiased_2011}: differences between the training and testing dataset distributions yield significant drops in performance. In order to address this problem, the field of domain adaptation (DA) proposes to build a transformation between the different distribution domains such that, when images are transformed from one {\it source} domain to other {\it target} domain, the distribution shift is reduced.

\begin{figure}[t]
\begin{center}
\small
\begin{tabularx}{.5\textwidth}{c|c}
Landsat-8 upscaled & Proba-V image \\
\midrule
2016-04-20 17:29 UTC & 2016-04-20 18:33 UTC\\
\includegraphics[width=4.2cm]{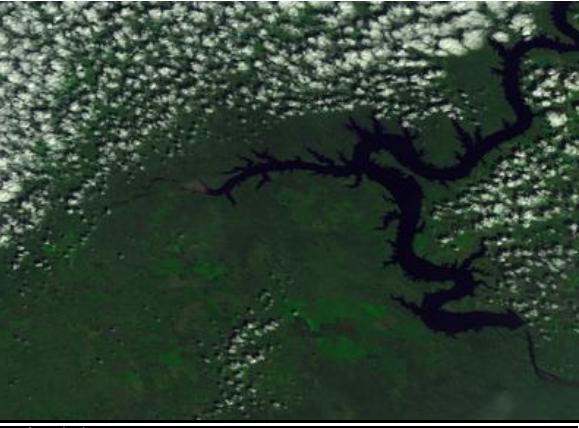} & 
\includegraphics[width=4.2cm]{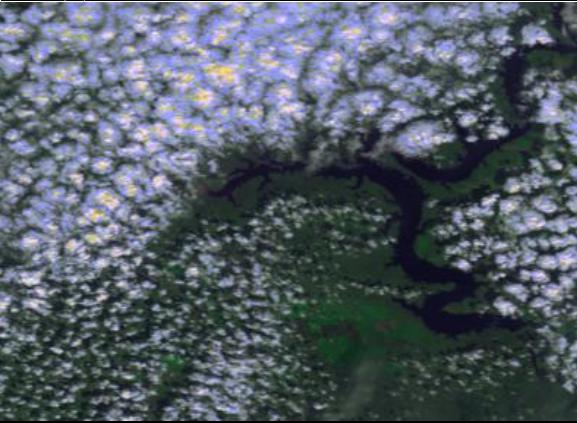} \\ 
2016-05-20 19:18 UTC & 2016-05-20 20:51 UTC\\
\includegraphics[width=4.2cm]{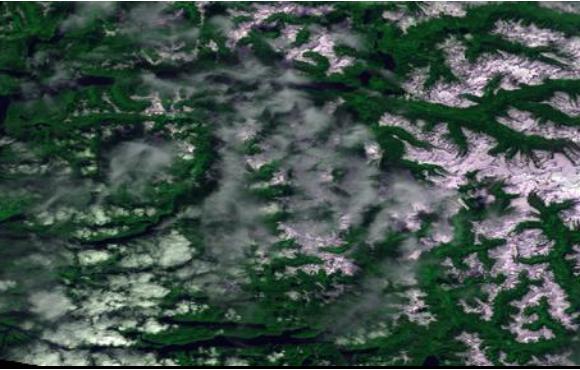} &
\includegraphics[width=4.2cm]{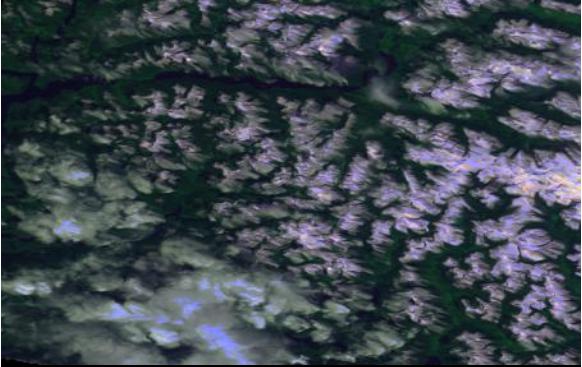} \\
2016-04-20 20:42 UTC & 2016-04-20 20:09 UTC\\
\includegraphics[width=4.2cm]{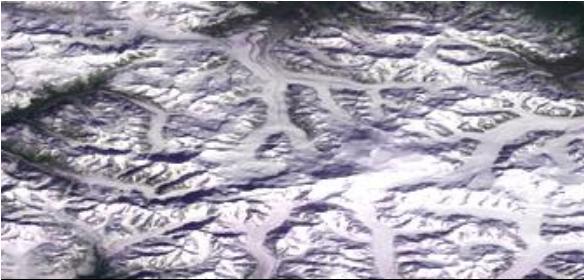} &
\includegraphics[width=4.2cm]{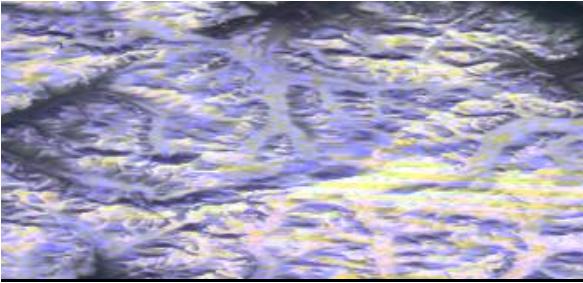} \\
\end{tabularx}
\end{center}
\caption{\small Close-in-time acquisitions of Landsat-8 and Proba-V satellites. Landsat-8 image is transformed and upscaled to resemble the optical characteristics of the Proba-V sensor; however, differences in radiometry and texture between images still remain. First row: Missouri river in North America (2016-04-20). Second and third row: North West Pacific coast, North America (2016-05-20 and 2016-04-20). 
\label{fig:pseudosimultaneouspvlandsat}}
\vspace*{-0.5cm}
\end{figure}

In this work we focus on the problem where the training ({\it source}) distribution corresponds to images and ground truth from one satellite, whereas the testing ({\it target}) distribution corresponds to images from another sensor. Notice that this a very broad scenario that is found frequently in Remote Sensing (RS). Examples of products built on this manner include cloud masks~\cite{mateo-garcia_domain_2019,mateo-garcia_transferring_2020}, but also land use and land cover classification~\cite{Banerjee17JSTARS}, vegetation indexes retrieval~\cite{svendsen_joint_2019}, or crop yield estimation~\cite{wolanin_estimating_2020}. The goal of domain adaptation is thus to find a transformation that allows models working on a given satellite (source domain) to work accurately on another one (target domain).

As a representative case study, in this work, we focus on the Landsat-8~\cite{Landsat8handbook} and Proba-V~\cite{Dierckx14} satellites. 
Transfer learning across these two sensors is particularly interesting since Landsat is a pioneering RS satellite program with a strong and well-established community and, hence, a good number of manually annotated datasets with cloud masks are publicly available, which could be very valuable to develop cloud detection models for Proba-V. Nevertheless, in order to build a model for Proba-V using Landsat-8 training data, differences in the imaging instruments on-board Landsat-8 and Proba-V must be taken in to account. On the one hand, the Operational Land Imager instrument (OLI), on board of Landsat-8, measures radiance in 9 bands in the visible and infrared part of the electromagnetic spectrum at 30m resolution. On the other hand, the Proba-V instrument retrieves 4 bands in the blue, red, near-infrared and short-wave infrared at 333m resolution  (full swath). Compared to Landsat-8, Proba-V has a much wider swath, which yields a more frequent revisiting time of around two days, whereas Landsat-8 revisit time ranges between 7 to 14 days. Figure~\ref{fig:pseudosimultaneouspvlandsat} shows three calibrated top of atmosphere (TOA) reflectance Landsat-8 and Proba-V images from the same location acquired with a difference of 30 to 90 minutes. We can see that, despite showing the same bands for Landsat-8 and Proba-V, and upscaling the Landsat-8 image to the Proba-V resolution (cf. section \ref{sec:pbda}), differences between the images still remain~\cite{sterckx_radiometric_2019}. 
In particular, Proba-V images are more blueish, due to saturation effects in the blue channel, and are more noisy~\cite{Sterckx13}. This higher noise and lower spatial resolution can be appreciated for example in the bottom row of Fig.~\ref{fig:pseudosimultaneouspvlandsat}, which shows a mountainous area in British Columbia where details in the Landsat-8 image are sharper than in the Proba-V one. Hence, products built using data in the Landsat-8 domain, such as the cloud detection model proposed in~\cite{mateo-garcia_transferring_2020}, show a drop in detection accuracy when directly applied to Proba-V images.  

In this work, we propose a DA methodology based on the \emph{state-of-the-art} work in DA for computer vision of Hoffman et al., CyCADA~\cite{hoffman_cycada_2018}, to the remote sensing field. In particular, we propose to use cycle consistent generative adversarial networks (cycleGAN~\cite{CycleGAN2017}) to find a DA transformation from the Proba-V domain to the Landsat-8 upscaled domain that removes noise and saturation of Proba-V images while preventing artifacts on the resulting images. One of the main advantages of the proposed methodology is that it is unpaired, i.e. it does not require a paired dataset of simultaneous and collocated Landsat-8 and Proba-V images to be trained. This is crucial for applications such as cloud detection since clouds presence and location highly varies between acquisitions. This can also be viewed in Fig.~\ref{fig:pseudosimultaneouspvlandsat}, even though acquisitions are the closest possible in time for Landsat-8 and Proba-V, cloud location changes significantly between the acquisitions. This would make a paired approach unfeasible. Following the proposed methodology, Proba-V images are enhanced and are shown to be statistically more similar to the Landsat-8 upscaled ones. 
In addition, from a transfer learning perspective, it is important to remark that the cloud detection models applied to Proba-V are trained using only Landsat-8 images and their ground truth.
In this context, a boost in cloud detection accuracy is shown for Proba-V when the proposed adversarial domain adaptation transformation is applied. 

The rest of the paper is organized as follows: in section~\ref{sec:relwork}, we discuss related work in cloud detection and domain adaptation in RS; in section~\ref{sec:method}, we detailed the proposed methodology to upscale Landsat-8 images and to train the domain adaptation network; section~\ref{sec:datasets} describes the Proba-V and Landsat-8 datasets where experiments are carried out. Finally, sections~\ref{sec:results} and \ref{sec:conclusions} contain the experimental results and discussion and conclusions, respectively.

\section{Related work}\label{sec:relwork}

There is a huge amount of work in both cloud detection and domain adaptation in the remote sensing literature. We discuss related work in both fields with a particular focus on approaches that deal with data coming from different sensors.
 
 \subsection{Transfer learning for Cloud detection}\label{sec:relworkclouds}
 
 Cloud detection has been lately dominated by deep learning approaches where, given a sufficiently large corpus of manually annotated images with the corresponding ground truth cloud masks, a network based on spatial convolutions is trained by back-propagation. Fully convolutional neural networks (FCNN)~\cite{long_fully_2015}, most of them based on the U-Net architecture~\cite{ronneberger_u-net_2015}, produce very accurate results and have the advantage that they can be applied to images of arbitrary size with a fast inference time. Works such as Jeppesen et al.~\cite{RSNet}, 
 Mohajerani and Sahedi~\cite{38-cloud-1,38-cloud-2}, Li et al.~\cite{li_deep_2019}, or Yang et al.~\cite{yang_cdnet:_2019} tackle cloud detection in Landsat-8 using Fully Convolutional Neural Networks trained in publicly available manually annotated datasets. They all show very high cloud detection accuracy outperforming the operational Landsat-8 cloud detection algorithm, FMask~\cite{zhu_improvement_2015}. Hence, our work seeks to transfer those accurate cloud detection models to other satellite data with a minimal drop in performance. There are some very recent works that propose to transfer a FCNN cloud detection model between different sensors. For instance, in Wieland et al.~\cite{wieland_multi-sensor_2019}, a FCNN is trained with contrast and brightness data augmentation in the Landsat-8 \textit{SPARCS} dataset~\cite{geological_survey_l8_2016} and it is tested on Sentinel-2, Landsat-7 and Landsat-8 images. Results show similar performance of the model on the three sensors; which suggest that the Sentinel-2 and the Landsat sensors are very similar and thus the data shift problem is not so relevant. On the other hand, in the work of Segal et al.~\cite{segal-rozenhaimer_cloud_2020}, a FCNN is trained on a manually annotated collection of World-View-2 images over the Fiji islands and tested in both World-View-2 and Sentinel-2 imagery. In this case, a significant drop in performance is observed in the Sentinel-2 domain; in order to correct this gap, authors propose a simple domain adversarial method which obtain good results but it is still far from the accuracy obtained in Word-View-2 data. The work of Shendryk et al.~\cite{shendryk_deep_2019} also proposes transfer learning, in this case using PlanetScope and Sentinel-2 imagery. The proposed network classifies patches as cloudy or clear instead of providing a full segmentation mask at pixel level. Nevertheless, an small gap in performance is observed between results in the source domain (PlanetScope) and the target domain (Sentinel-2). Finally, in our previous work~\cite{mateo-garcia_transferring_2020}, we showed that transfer learning from Proba-V to Landsat-8 and from Landsat-8 to Proba-V produce accurate results on a par with the FMask~\cite{zhu_improvement_2015} model on Landsat-8 and surpassing the operational cloud detection model~\cite{tote_evaluation_2018} for Proba-V, respectively. However, a significant gap between transfer learning approaches and \emph{state-of-the-art} models trained with data from the same domain still exists, which is the focus of the present work. 
 
 \subsection{Domain adaptation}
 
 The remote sensing community has traditionally addressed DA taking advantage of a deep understanding of the imaging sensors and exploiting their physical and optical characteristics to provide well-calibrated products~\cite{QRSLiangBook}. Despite the efforts to provide a good calibration, small differences are found in retrieved radiance values due to different spectral response functions, saturation effects, mixed pixels, etc~\cite{mandanici_preliminary_2016,revel_sentinel-2a_2019,sterckx_radiometric_2019}. To this end several works propose sensor inter-calibrations or harmonizations to correct biases between the bands of different sensors~\cite{Zhao2018JSTARS,houborg_cubesat_2018,claverie_harmonized_2018,zhang_characterization_2018,scheffler_spectral_2020}. These approaches train models in \emph{paired} data using either real spectra of both satellites (retrieved in same location and close in time) or simulated radiances. As mentioned earlier, \emph{paired} approaches cannot be applied to cloud detection due to the extreme variability of cloud location between acquisitions. Among \emph{unpaired} approaches histogram matching~\cite{digitalimproc} aligns the observed radiance distribution of both satellites using the cumulative density function of the data. Histogram matching is fast and reliable and thus we use it as a baseline to compare our method. More complex \emph{unpaired} methods include multivariate matching~\cite{inamdar_multidimensional_2008}, graph matching~\cite{tuia_graph_2013} or manifold alignment~\cite{tuia_semisupervised_2014}.
 
 All the methods discussed so far only focus on the spectral information of images disregarding the spatial dimension. In order to account for spatial changes, recent works propose DA using convolutional neural networks (CNN). Most of these works use Generative Adversarial Networks (GAN)~\cite{goodfellow_2014} to align source and target distributions. Those works could be divided in \emph{feature-level} DA and \emph{pixel-level} DA. In \emph{feature-level} or \emph{discriminative} DA~\cite{ganin_domain-adversarial_2016,tzeng_adversarial_2017} the model to be transferred is jointly trained with its normal loss and to make its internal representations (i.e. activations at some layers) invariant to the input distribution. Hence, \emph{feature level} DA requires retraining the model with that extra penalty. An example of \emph{feature level DA} is the work of Segal et al.~\cite{segal-rozenhaimer_cloud_2020} previously discussed in section~\ref{sec:relworkclouds}. In \emph{pixel-level} DA~\cite{liu_unsupervised_2017,bousmalis_unsupervised_2017} (also called \emph{image-to-image} DA), extra networks are trained to transform images between domains. Hence. \emph{pixel-level} DA is independent of the transferred model and thus it could be applied to other problems with same inputs. Our work falls into the \emph{pixel-level} DA framework: we assume that the cloud detection model trained in the source domain is fixed and thus we focus on finding a DA transformation from the target to the source domain (section~\ref{sec:method}).
 
 Regarding the definition and the types of domains in remote sensing, works such as~\cite{elshamli_domain_2017,song_domain_2019,tasar_colormapgan_2020,koga_method_2020} consider data from a single sensor, where the source and target domains are represented by images from different locations or different time acquisitions. DA works where domains are represented by different sensors are scarce; for instance, the work of Benjdira et al.~\cite{benjdira_unsupervised_2019} tackles urban segmentation in aerial imagery of two cities acquired with two different cameras. They obtain good results despite differences in spectral bands and spatial resolution of the instruments are not taken into account. Even though there are some works using GANs to transfer learning between different sensors, most of them involve training the classifiers using some labeled samples from the target domain. This is the case of works that tackle DA between SAR and optical images~\cite{ye_sar_2019,Wang19_J-GRSL_8825802,Liu18_C-IGARSS_8517866}. It is important to remark that we are dealing with \emph{unsupervised domain adaptation}~\cite{DARSTuia2016,ganin_domain-adversarial_2016} (also known as \emph{transductive transfer learning}~\cite{pan_survey_2010}), which assumes there is no labeled data available in the target domain.

\begin{figure}[t]
\begin{center}
\includegraphics[width=8.5cm]{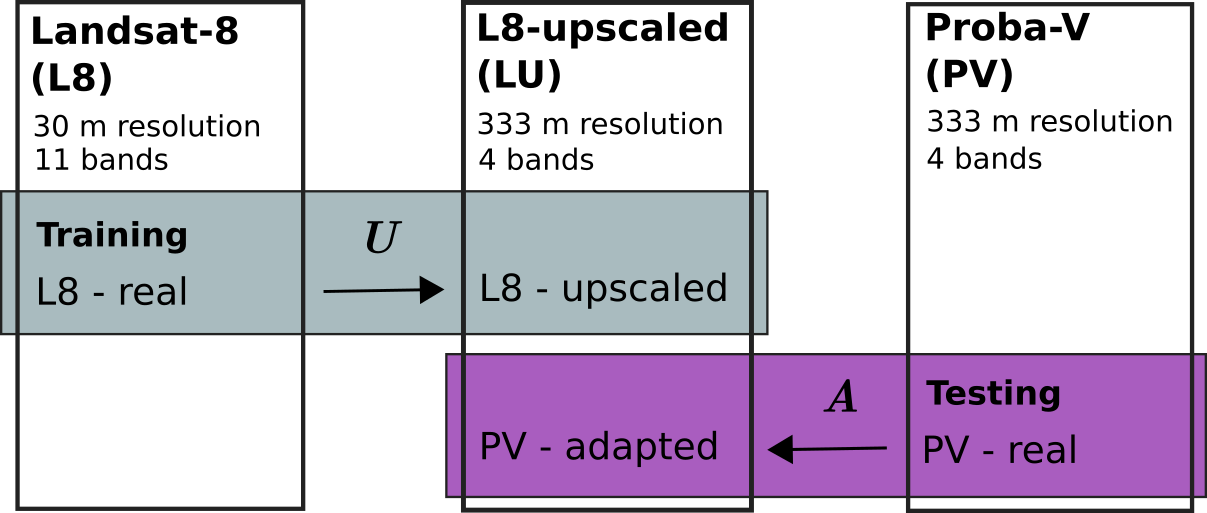} 
\end{center}
\caption{\small Transfer learning and adaptation scheme: Landsat-8 and Proba-V datasets and how they are transformed between the three different domains. The transformations look for adaptation between the domains: $U$ is the upscaling transformation applied to Landsat-8 to resemble the Proba-V instrument characteristics (sec.~\ref{sec:pbda}); and $A$ adapts from the Proba-V domain to the Landsat-8 upscaled domain (sec.~\ref{sec:gada}). 
\label{fig:methodology}}
\vspace*{-0.5cm}
\end{figure}

\section{Methodology}\label{sec:method}

We assume two independent datasets from two different sensors are given, but we only have labels for one dataset. The main idea is to be able to use the data from the labeled dataset in order to design algorithms to solve problems in the unlabeled dataset. 

In our particular case, we have images for Landsat-8 (L8) with the corresponding ground truth cloud masks (binary labels identifying \emph{cloudy} or \emph{clear} pixels), $\{X_{\LA},y_{\LA}\}$; and we only have Proba-V (PV) images without ground truth, $X_{\PV}$. 
Therefore, we want to perform cloud detection in $X_{\PV}$ using algorithms trained with $\{X_{\LA},y_{\LA}\}$. Since we know the technical specifications of Landsat-8 and Proba-V, we can design an upscaling algorithm to convert images captured from Landsat-8 to resemble Proba-V spectral and spatial characteristics. This upscaling could work quite well, and actually classical remote sensing approaches follow this methodology to combine or perform transfer learning across different existing satellites. 
However, this upscaling transformation is not perfect since it is based on the pre-launch characterization of the instruments and is always susceptible to be affected by diverse uncertainty sources. Therefore, an extra adaptation step could be used in order to transform the Proba-V images before applying the transfer learning algorithms. In this work, we are going to explore how to design this extra step by using Generative Adversarial Networks (GANs).

\begin{figure*}[ht]
    \centering
    \includegraphics[width=.99\linewidth]{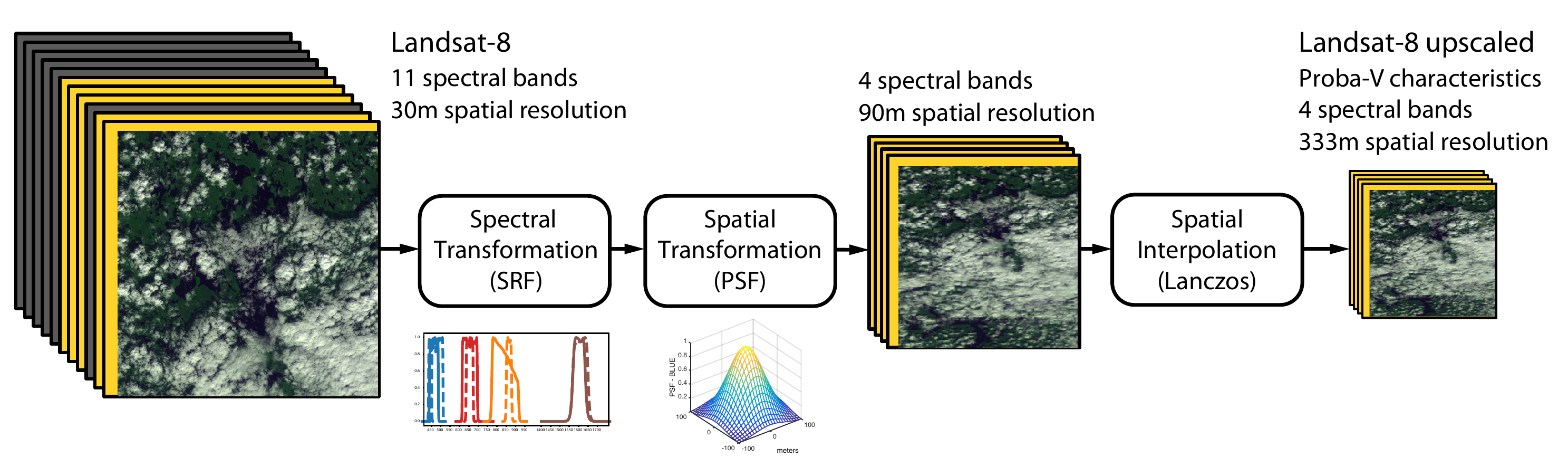}\\
    \caption{Upscaling transformation ($U$ in Fig.~\ref{fig:methodology}) applied to Landsat-8 in order to resemble the Proba-V instrument characteristics.
    }
    \label{fig:ilustrationtransform}
\end{figure*}

In Fig.~\ref{fig:methodology}, we show the proposed adaptation scheme. The upscaling transformation, $U$, converts the Landsat-8 labeled data to a domain where it has similar spatio-spectral properties than Proba-V (i.e. the same number of bands and same spatial resolution). We can use the Landsat-8 upscaled (LU) data in order to train an algorithm, in our case we are going to design a cloud detection algorithm using fully convolutional neural networks. While this model could be applied directly on Proba-V images, we will show that an extra adaptation step, $A$, applied to the Proba-V images to resemble even more the upscaled domain, could improve the similarity between the training and the testing data and therefore improve the performance of the cloud detection algorithm. Note that, if no adaptation is used, $A$ would be the identity function. In the following subsections, we detail both the transformation $U$, based on the instruments characteristics, and the transformation $A$, which is based on CycleGANs.  

\subsection{Upscaling transformation from Landsat-8 to Proba-V} \label{sec:pbda}

\begin{figure}[ht]
    \centering
    \includegraphics[width=.9\linewidth]{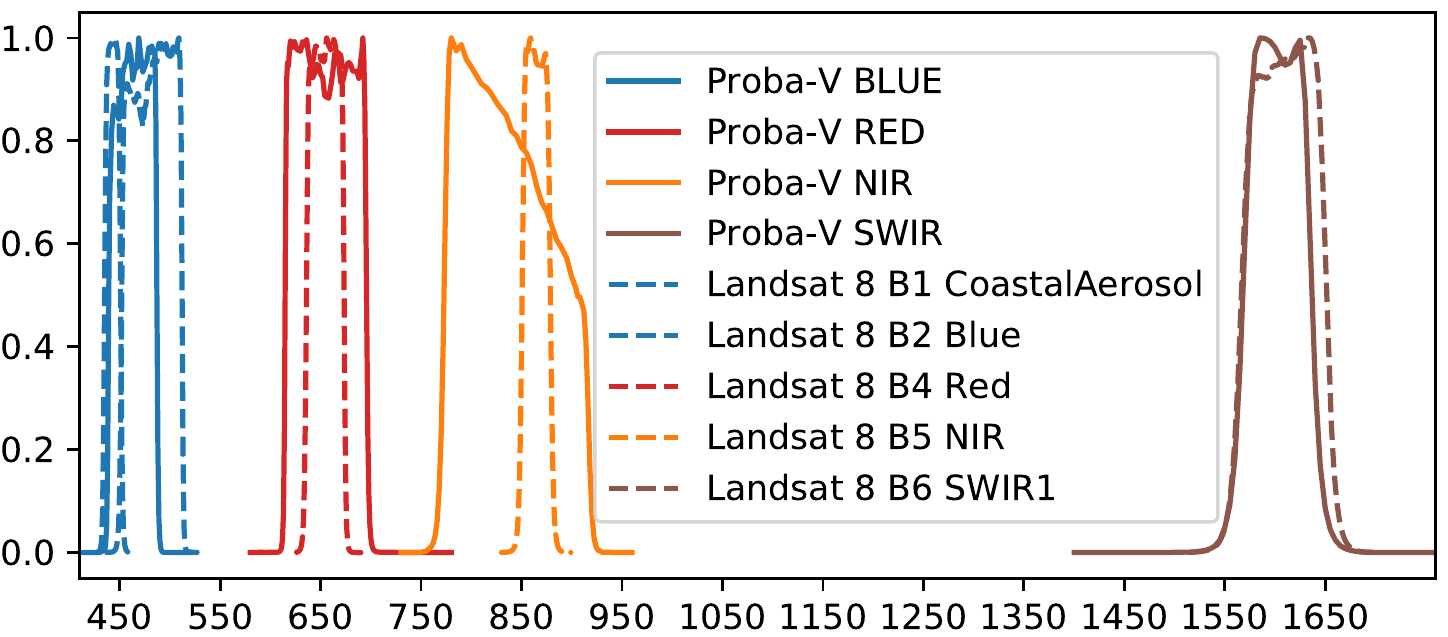}
    \caption{Spectral response of Landsat-8 and Proba-V channels.}
    \label{fig:spectralresponseall}
\end{figure}

In this section, we describe the upscaling transformation ($U$ transformation in Fig.~ \ref{fig:methodology}). 
In a standard transfer learning approach from one sensor to another, the first step is to transform Landsat-8 images to resemble Proba-V images in terms of spectral and spatial properties.
Figure~\ref{fig:ilustrationtransform} shows the proposed upscaling from Landsat-8 to Proba-V using the instrumental characteristics of both sensors. First, we select the spectral bands from Landsat-8 that overlap with Proba-V in terms of the spectral response functions (SRF) of both instruments. 
Figure~\ref{fig:spectralresponseall} shows the SRF of overlapping bands in Landsat-8 (dashed) and Proba-V (solid). SWIR and RED bands present the best agreement. However, the NIR SRF of Proba-V is wider and its peak is not aligned with Landsat-8 B5 band, which might led to differences in the the retrieved radiance. Finally, the BLUE band of Proba-V overlaps with two different Landsat-8 bands. Therefore, the contribution of Landsat-8 B1 and B2 bands is weighted according to the overlapping area of the SRFs: 25\% and 75\% for B1 and B2, respectively.

Then, the selected bands of the Landsat-8 image are scaled to match the spatial properties of Proba-V.
The 30m resolution Landsat-8 bands are upscaled to the 333m resolution of Proba-V. This upscaling takes into account the optical characteristics of the Proba-V sensor and the re-sampling of the 333m product described in~\cite{Dierckx14}.
First, the point spread function (PSF) of each Proba-V spectral band is used to convert the Landsat-8 observations to the nominal Proba-V spatial resolution at nadir. 
The ground sampling distance (GSD) for the Proba-V center camera is about 96.9m for the BLUE, RED and NIR channels, while the SWIR center camera resolution is 184.7m~\cite{Dierckx14} 
The SWIR PSF is about twice as wide as the PSF of the other bands, which stresses the fact that a distinct spatial adaptation might be applied to each band. 
The PSFs of the bands are modeled as 2 dimensional Gaussian filters, which are applied to the 30m resolution Landsat-8 bands. The filtered image is upscaled to the nominal 90m resolution at nadir by taking 1 out of every 3 pixels. Finally, Lanczos interpolation is applied to upscale the image to the final 333m Proba-V resolution. Notice that Lanczos is the interpolation method used at the Proba-V ground segment processing to upscale the acquired raw Proba-V data to the 333m Plate Carée grid~\cite{Dierckx14}.
Ground truth labels, $y_L$ at 30m, must were also scaled to get a Landsat-8 upscaled dataset at 333m: $\{X_{\LU},y_{\LU}\}$. 


\subsection{Transfer Cloud Detection Model} \label{sec:fcnn}

The cloud detection model trained on the Landsat-8 upscaled dataset is a FCNN classifier based on the simplified U-Net architecture described in~\cite{mateo-garcia_transferring_2020}. This model is trained in the Landsat-8 Upscaled dataset ($\{X_{\LU},y_{\LU}\}$); hence, it takes as input a 4-band 333m resolution image and produces a binary cloud mask. Therefore, it could be applied directly to Proba-V images. Nevertheless, as explained before, statistical differences between Landsat-8 upscaled and Proba-V images make that the performance of this model is not as good as expected. This effect is related to the different sensor spectral response functions, saturation effects, radiometric calibration, modulation transfer functions, or mixed pixels. For instance, as shown in Fig.~\ref{fig:pseudosimultaneouspvlandsat}, Proba-V contains many saturated pixels, specially in the blue channel, which is a known issue. This suggests that an extra domain adaptation step could be added to improve the transfer learning results reported in~\cite{mateo-garcia_transferring_2020}.



\subsection{Generative Adversarial Domain Adaptation} \label{sec:gada}

\begin{figure*}[ht]
    \centering
    \includegraphics[width=1\linewidth]{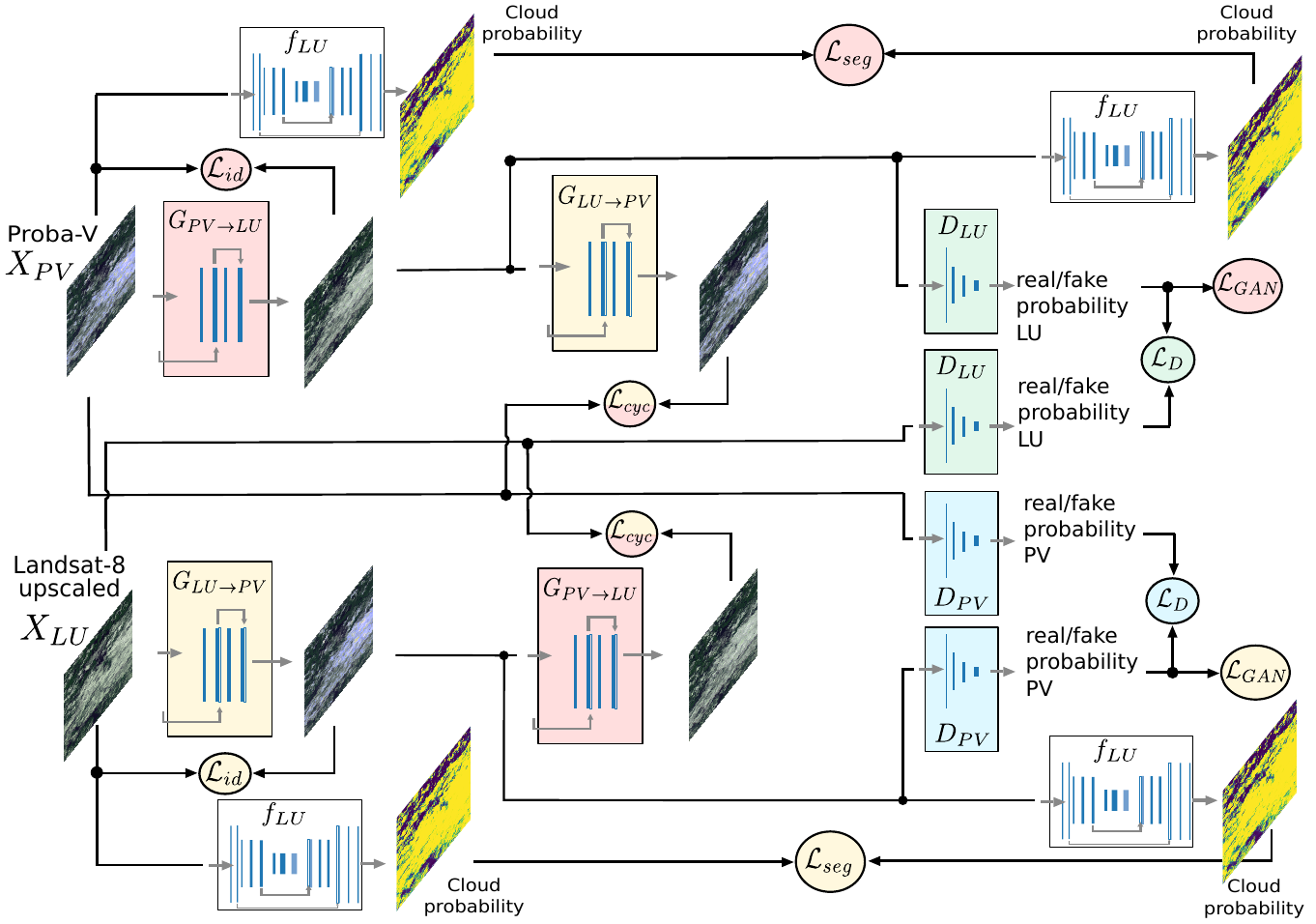}\\
    \caption{Scheme of the forward passes for the training procedure of the proposed cycle consistent adversarial domain adaptation method. The four networks ($G_{\PV\to\LU}$, $G_{\LU\to\PV}$,  $D_{\PV}$, $D_{\LU}$) have a different color. Losses are depicted with circles and their fill color corresponds to the color of the network that they penalize.}
    \label{fig:ganframework}
\end{figure*}

In this section, we describe the training process for the extra adaptation transformation ($A$ in Fig.~\ref{fig:methodology}) that we propose to improve the performance of the transferred models. This training process is based on the Generative Adversarial Network (GAN)~\cite{goodfellow_2014} framework. 


The main idea of GANs is to train two networks, a generative one and a discriminative one, with opposite objectives, simultaneously. This adversarial training fits a data generator that minimizes the Jensen-Shannon divergence between the real and the generated data distribution. An extension of the original GANs formulation, the conditional GANs~\cite{isola2017image}, was proposed to train a model that generates samples from a conditional distribution. One application of conditional GANs is the Generative Adversarial Domain Adaptation (GADA) proposed in ~\cite{tzeng_adversarial_2017,bousmalis_unsupervised_2017,hoffman_cycada_2018}. In those works, the conditional GANs formulation was modified to solve domain adaptation problems.

Probably, the most complete approximation for Domain adaptation based on GANs is the one proposed in CyCADA~\cite{hoffman_cycada_2018}. Unlike the classical GANs, where adaptation is performed in one direction only, this approach proposes a double simultaneous adaptation between the two domains. This allows to include several consistency terms in order to impose restrictions on the two adaptation directions.    
Our approach has a similar structure than CyCADA (Fig.~\ref{fig:ganframework}). It has two Generators and two Discriminators: $G_{\LU\to \PV}$, $G_{\PV\to \LU}$, $D_{\PV}$, $D_{\LU}$. We are interested in using $G_{\PV\to \LU}$ to adapt the Proba-V images to better match the upscaled ones that have been used to train the cloud detection algorithm, i.e. $A \equiv G_{\PV\to \LU}$.

On the one hand, the discriminators are trained to minimize the binary cross-entropy loss between the real and the generated images:
$$
\begin{aligned}
\mathcal{L}_D(D_{\LU}) &= \sum_i -\log(D(X^i_{\LU})) + \\ &\quad \quad - \log(1- D(G_{\PV\to \LU}(X^i_{\PV}))) \\
\mathcal{L}_D(D_{\PV}) &= \sum_i -\log(D(X^i_{\PV}))  + \\ &\quad \quad - \log(1- D(G_{\LU\to \PV}(X^i_{\LU})))
\end{aligned}
$$

On the other hand, the generators are trained to fool the discriminators by minimizing the adversarial loss:
$$
\begin{aligned}
\mathcal{L}_{GAN}(G_{\PV\to \LU}) &= \sum_i -\log(D(G_{\PV\to \LU}(X^i_{\PV}))) \\
\mathcal{L}_{GAN}(G_{\LU\to \PV}) &= \sum_i  - \log(D(G_{\LU\to \PV}(X^i_{\LU}))) 
\end{aligned}
$$

In this work, in order to ensure consistency between the real and the generated images, three extra penalties are added to the standard GAN generator loss: the {\it identity consistency loss}, the {\it cycle loss}, and the {\it segmentation consistency loss}. Firstly, the {\it identity consistency loss}, introduced in our previous work~\cite{mateo-garcia_domain_2019}, is added to make the TOA reflectance of the input similar to those in the output:
$$
\begin{aligned}
\mathcal{L}_{id}(G_{\PV\to \LU}) &= \sum_i \| X^i_{\PV} - G_{\PV\to \LU}(X^i_{\PV}) \|_1 \\
\mathcal{L}_{id}(G_{\LU\to \PV}) &= \sum_i  \| X^i_{\LU} - G_{\LU\to \PV}(X^i_{\LU})\|_1
\end{aligned}
$$
Secondly, {\it the cycle consistency loss}, proposed in~\cite{isola2017image}, is added to both generators to force them to act approximately as inverse functions one of each other:
$$
\begin{aligned}
\mathcal{L}_{cyc} & (G_{\PV\to \LU},G_{\LU\to \PV}) =  \\
&  \sum_i  \| X^i_{\PV} - G_{\LU\to \PV}(G_{\PV\to \LU}(X^i_{\PV})) \|_1 \\
+ & \sum_i \| X^i_{\LU} - G_{\PV\to \LU}(G_{\LU\to \PV}(X^i_{\LU})) \|_1
\end{aligned}
$$
Finally, we additionally include a {\it segmentation consistency loss} that takes advantage of the cloud detection model trained in the LU domain. We apply this model to both LU and PV images even though the cloud detection classifier ($f_{\LU}$) is trained using only LU images, the idea is that the LU classifier can act as a rough supervisor in the Proba-V domain. This approach is also taken in CyCADA~\cite{hoffman_cycada_2018}. The selected semantic segmentation loss is the Kullback-Leibler divergence between the cloud probabilities of the model $f_{\LU}$ applied to real and generated images:
$$
\begin{aligned}
\mathcal{L}_{seg}(G_{\PV\to \LU}) &=  \sum_i  {\rm KL}(f_{\LU}(X^i_{\PV}) | f_{\LU}(G_{\PV\to \LU}(X^i_{\PV})) ) \\
\mathcal{L}_{seg}(G_{\LU \to \PV}) &=  \sum_i {\rm KL}(f_{\LU}(X^i_{\LU}) | f_{\LU}(G_{\LU\to \PV}(X^i_{\LU})) ) \\
\end{aligned}
$$

Therefore, the final loss of the generators is the weighted sum of the five described losses:
$$
\begin{aligned}
\mathcal{L}(G_{\PV\to \LU}) &= \lambda_{GAN}\mathcal{L}_{GAN}(G_{\PV\to \LU}) + \lambda_{id}\mathcal{L}_{id}(G_{\PV\to \LU}) \\ 
&\quad + \lambda_{cyc}\mathcal{L}_{cyc}(G_{\PV\to \LU}, G_{\LU\to \PV}) \\ 
&\quad + \lambda_{seg}\mathcal{L}_{seg}(G_{\PV\to \LU})   \\
\mathcal{L}(G_{\LU\to \PV}) &= \lambda_{GAN}\mathcal{L}_{GAN}(G_{\LU\to \PV}) + \lambda_{id}\mathcal{L}_{id}(G_{\LU\to \PV}) \\ 
&\quad + \lambda_{cyc}\mathcal{L}_{cyc}(G_{\PV\to \LU}, G_{\LU\to \PV}) \\ 
&\quad + \lambda_{seg}\mathcal{L}_{seg}(G_{\LU\to \PV})
\end{aligned}
$$
The weight parameters are set to $\lambda_{cyc}=\lambda_{id}=5$ and  $\lambda_{seg}=\lambda_{GAN}=1$, so that losses are of the same magnitude.  
In addition, the two discriminators are regularized using a 0-centered gradient penalty with a weight of 10~\cite{Mescheder2018ICML}. In section~\ref{sec:results}, we conduct several experiments by setting some of these weights to zero in order to quantify the importance of each of these terms. 
In addition, notice that by setting some of these hyper-parameters to zero we obtain different adversarial domain adaptations proposals in the literature. In particular, 
if we set $\lambda_{id}=0$ we get the original CyCADA of Hoffman et al.~ \cite{hoffman_cycada_2018}. 
By setting $\lambda_{cyc}=\lambda_{seg}=0$ we get the approach of our previous work~\cite{mateo-garcia_domain_2019}. 
Finally, by setting $\lambda_{cyc}=\lambda_{seg}=\lambda_{id}=0$, we get the classical GAN approach.
Details about the training procedure and particular network architectures of the generators $G$ (fully convolutional neural networks) and discriminators $D$ (convolutional neural networks) of the proposed adaptation model can be found in Appendix~\ref{sec:gandetails}. 
Additionally, the implemented code is available at \url{https://github.com/IPL-UV/pvl8dagans}. 



From both a methodological and an operational perspective, the proposed approach has an important benefit: it does not require simultaneous and collocated pairs of Landsat-8, $X^i_{\LU}$, and Proba-V, $X_{\PV}^i$, images. Having coincident pairs from sensors on different platforms would be impossible in our case. Note that clouds presence and location within an image highly vary even for small time differences. This problem prevents the use of other approaches such as Canonical Correlation Analysis~\cite{Denaro20JSTARS} or directly learning a generic transformation from $X^i_{\PV}$ to $X^i_{\LU}$~\cite{Zhao2018JSTARS}.

\section{Manually annotated Datasets} \label{sec:datasets}

\begin{figure}
    \centering
    \includegraphics[width=\linewidth]{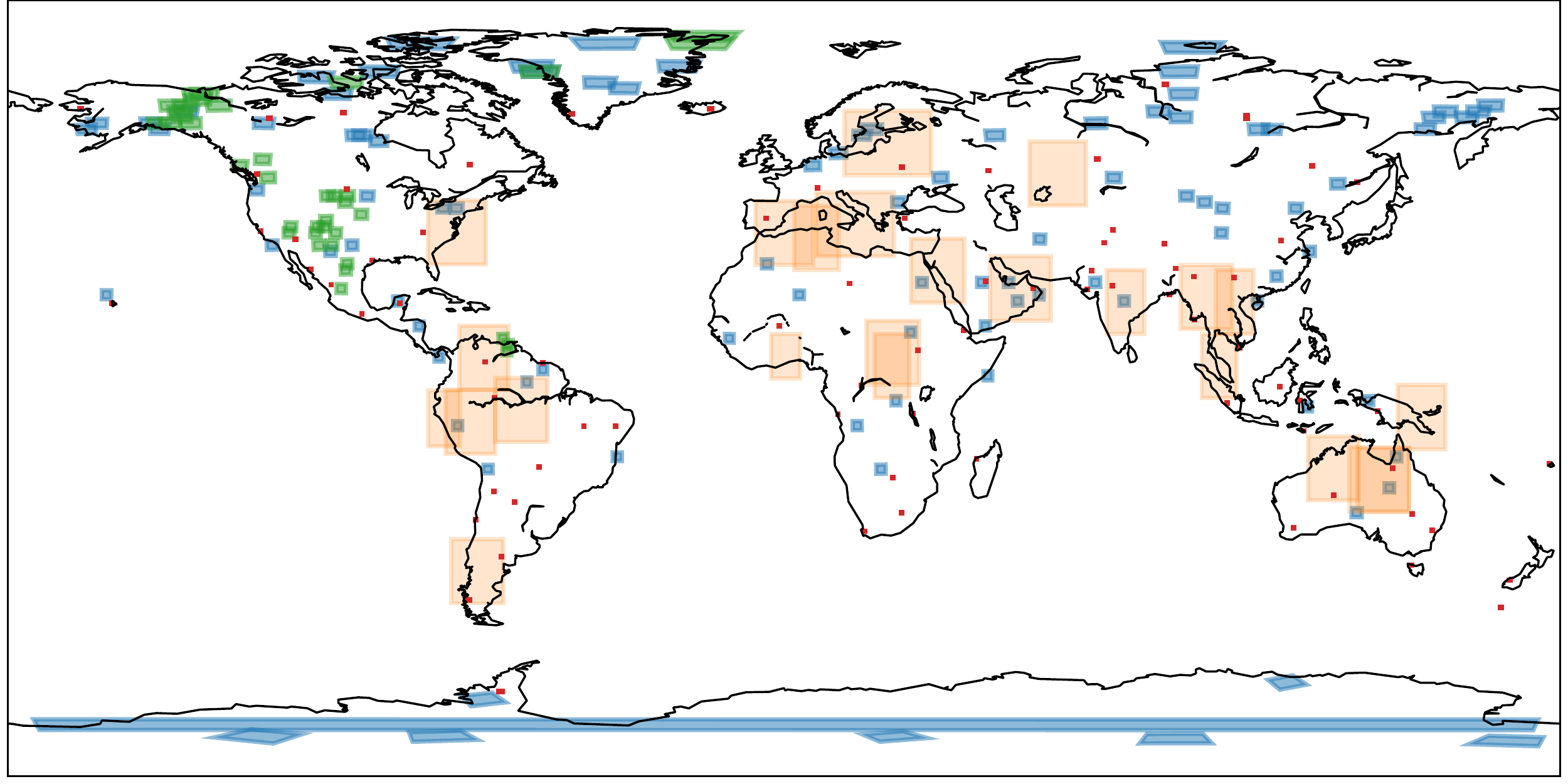}\\
    \textcolor[rgb]{1,0.5,0.05}{\bf $\blacksquare$} PV24~\cite{mateo-garcia_transferring_2020}
    \textcolor[rgb]{0.12156863,0.46666667,0.70588235}{\bf $\blacksquare$} Biome~\cite{geological_survey_l8_2016-1} 
    \textcolor[rgb]{.173, .627, .173}{\bf $\blacksquare$} 38-Clouds~\cite{38-cloud-1}
    \textcolor[rgb]{.839,.153,.157}{\bf $\blacksquare$}~SPARCS~\cite{geological_survey_l8_2016}
    \caption{Location of Landsat-8 and Proba-V products with manually annotated ground truth cloud mask. For the \textit{Biome}~\cite{geological_survey_l8_2016-1} and \textit{38-Clouds} dataset~\cite{38-cloud-1} we additionally downloaded Proba-V images from the same location and acquisition time if available.}
    \label{fig:locs}
\end{figure}

Transfer learning from Landsat-8 to Proba-V is supported by the fact that there are several open access datasets with manually labeled clouds for the Landsat-8 mission. In this work, we use three of them that have a large coverage of acquisitions across different dates, latitudes and landscapes. The \textit{Biome} dataset~\cite{geological_survey_l8_2016-1}, released for the Landsat-8 validation study of Foga et. al~\cite{foga_cloud_2017}, is the largest among them. It consists of 96 full acquisitions covering the different biomes on Earth. The \textit{SPARCS} dataset, collected in the study of Hughes and Hayes~\cite{hughes_automated_2014}, contains 80 1,000$\times$1,000 patches from different Landsat-8 acquisitions. Finally, the \textit{38-Clouds} dataset of Mohajerani and Saeedi~\cite{38-cloud-1} has 38 full scenes mostly located in North America. Images and ground truth cloud masks from these datasets have been upscaled, to match Proba-V spectral and spatial properties, following the procedure described in section~\ref{sec:pbda}. Hence, when we refer to those datasets, we assume 4-band images and ground truth at 333m. 
For testing the proposed cloud detection approach in Proba-V, since there are not publicly available datasets, we use the \textit{PV24} dataset created by the authors in~\cite{gomez-chova_cloud_2017} and extensively curated in~\cite{mateo-garcia_transferring_2020}. The \textit{PV24} dataset contains 24 full Proba-V images at 333m resolution and their corresponding manually annotated cloud masks. Figure~\ref{fig:locs} shows the locations of the products of the aforementioned datasets. It is important to remark that we only use data from the \textit{Biome} dataset for training  the cloud detection models based on FCNN (section~\ref{sec:fcnn}); the other three datasets (\textit{SPARCS}, \textit{38-Cloud} and \textit{PV24}) are only used for testing the models.

On the other hand, to train the proposed domain adaptation model based on CycleGANs (section~\ref{sec:gada}), a set of 181 Proba-V products from the same locations and season as the \textit{Biome} dataset has been selected. Using those Proba-V images and the Landsat-8 upscaled images from the \textit{Biome} dataset, we created the \textit{Biome Proba-V pseudo-simultaneous dataset}, which contains 37,310 pairs of patches of 64$\times$64 pixels, used to train the proposed DA method. Notice that, in this dataset, the pairs of images, one coming from Proba-V and the other from Landsat-8, are images from the same location and close-in-time acquisitions when available\footnote{Some images from the \textit{Biome} dataset are previous to the beginning of the Proba-V mission catalog; in this cases, we use images from same day of year in the next year.}. The same approach is followed to create the \textit{38-Clouds Proba-V pseudo-simultaneous dataset}, which is only used for testing the domain adaptation results. 
Images of this dataset, together with the results of the proposed domain adaptation and cloud detection models, are available at \url{https://isp.uv.es/projects/cloudsat/pvl8dagans}.

Finally, it is important to point out that images in this work are operational level-1TP products for Landsat-8 and level-2A products for Proba-V. These products have been pre-processed to top of atmosphere (TOA) reflectance for both Landsat-8~\cite{Landsat8handbook} and Proba-V~\cite{Wolters18}. 

\section{Experimental Results} \label{sec:results}

This section is divided in two parts. In the first one, we analyze the radiometric and spatial properties of resulting images. The purpose is to assess the quality of the proposed transformation. We show that the proposed DA transformation produces reliable images (i.e. images without artifacts) that are statistically similar to Landsat-8 upscaled images. Additionally we show that pixels flagged as good radiometric quality in Proba-V are much less changed by the proposed DA transformation. In the second part, we analyze the impact of DA on the cloud detection performance. Cloud detection results in the source domain (Landsat-8 upscaled) are compared with results in the target domain (Proba-V), with and without the DA transformation. In addition, we conduct an ablation study to show the relative importance of each of the proposed loss terms included to fit the generator network. 

\begin{figure}[t]
    \centering
    \includegraphics[width=.7\linewidth]{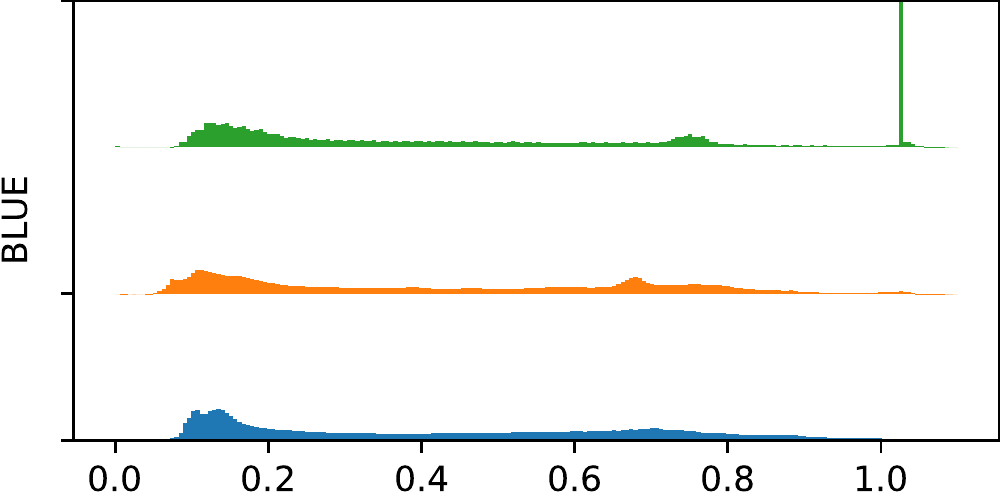}\\
    \includegraphics[width=.7\linewidth]{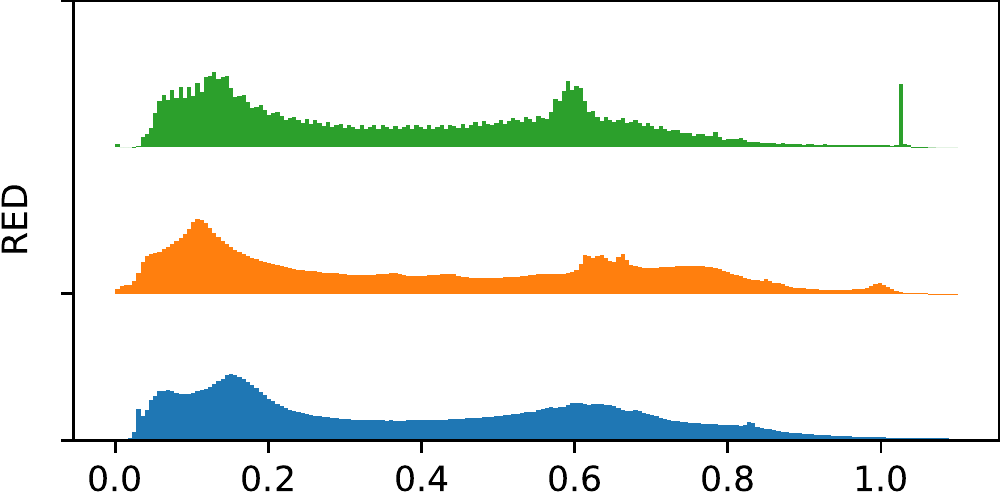}\\
    \includegraphics[width=.7\linewidth]{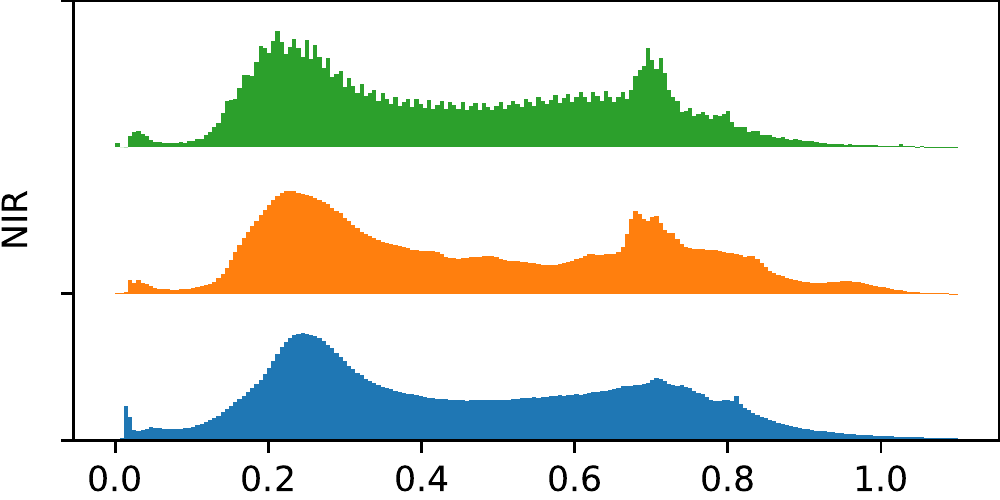}\\
    \includegraphics[width=.7\linewidth]{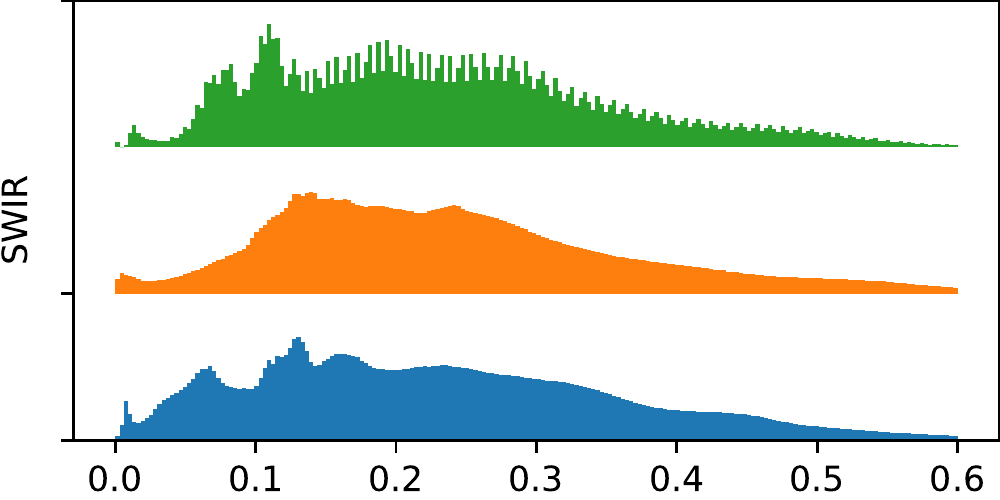}\\
    \includegraphics[width=.9\linewidth]{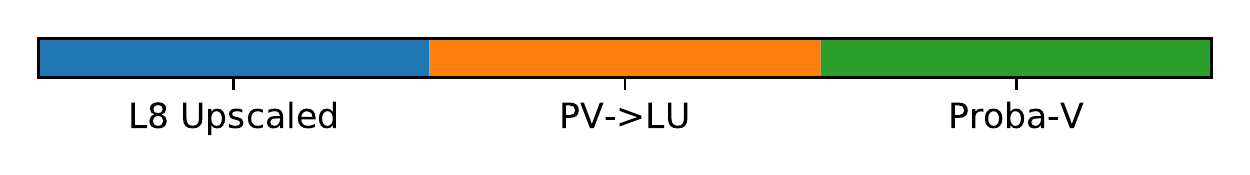} 
    \caption{TOA refectance distribution on each of the spectral bands for Proba-V (green) images, Proba-V images transformed using the proposed DA method (orange) and pseudo-simultaneous Landsat-8 Upscaled images (blue). Values measured across all image pairs in the \textit{38-Clouds pseudo-simultaneous dataset}}   \label{fig:histograms}
\end{figure}

\subsection{Domain Adaptation of input images}


As explained in section~\ref{sec:datasets}, we trained the DA method described in section~\ref{sec:gada} using Proba-V and Landsat-8 upscaled patches from the \textit{Biome Proba-V pseudo-simultaneous dataset}. The generator and discriminator networks are trained simultaneously with mini-batch stochastic gradient descent (see details in sec.~\ref{sec:gandetails}). Afterwards, the trained Proba-V to Landsat-8 upscaled generator $G_{\PV\to \LU}$ ($A$ in Fig.~\ref{fig:methodology}) is evaluated in the \textit{38-Clouds Proba-V pseudo-simultaneous dataset}, (i.e. the $G_{\PV\to \LU}$ network is applied to all Proba-V images in the dataset). Figure~\ref{fig:histograms} shows the distribution of TOA reflectance values for each of the bands without the domain adaptation step (green) and after applying $G_{\PV\to \LU}$ (orange) for all the Proba-V images in the dataset. 
One of the interesting results from these distributions is that the characteristic saturation in the Blue and Red bands of Proba-V disappears in the adapted images. In addition, the shape of the distribution of the adapted data is more similar to the shape of the pseudo-simultaneous Landsat-8 upscaled images (blue).

\begin{figure*}[t]
    \begin{center}
    \setlength{\tabcolsep}{1.5pt}
    \begin{tabularx}{\linewidth}{ccc}
    Proba-V (PV)& PV $\to$ LU & L8-Upscaled (LU)\\
    \includegraphics[width=.32\linewidth,height=5cm]{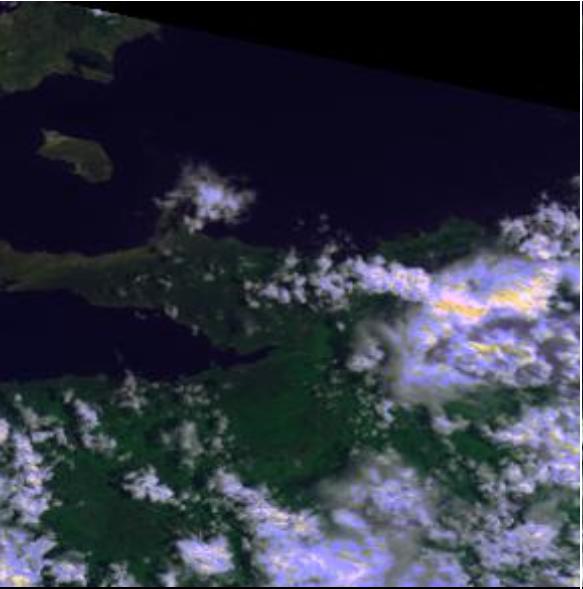} &
    \includegraphics[width=.32\linewidth,height=5cm]{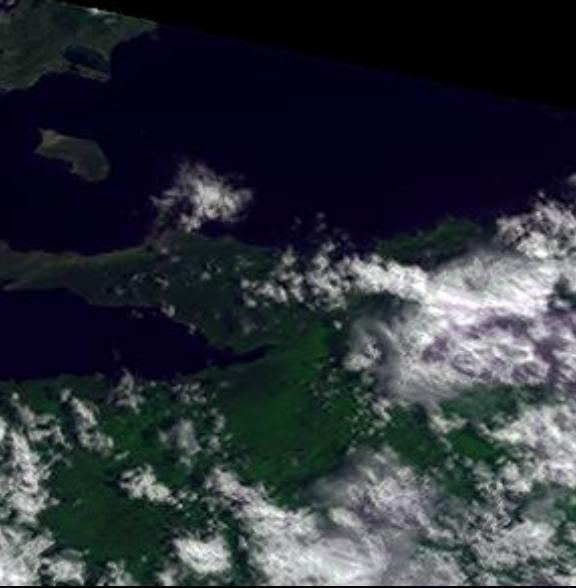} &
    \includegraphics[width=.32\linewidth,height=5cm]{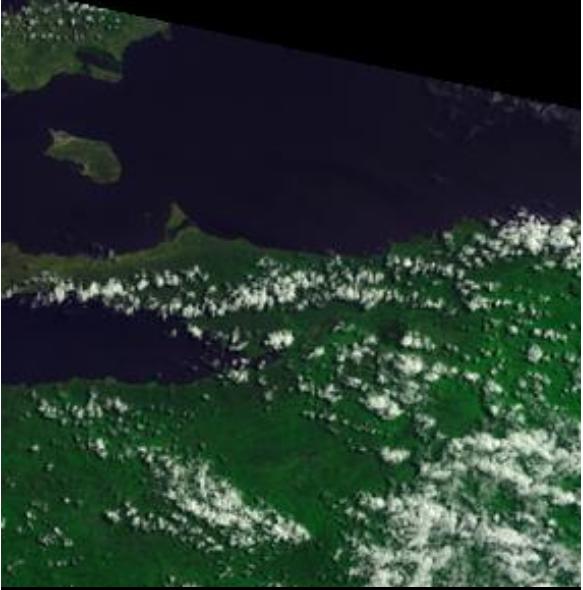}\\
    \includegraphics[width=.32\linewidth,height=5cm]{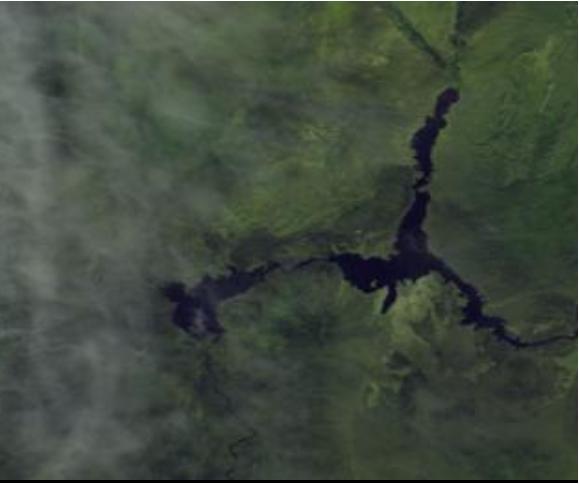} &
    \includegraphics[width=.32\linewidth,height=5cm]{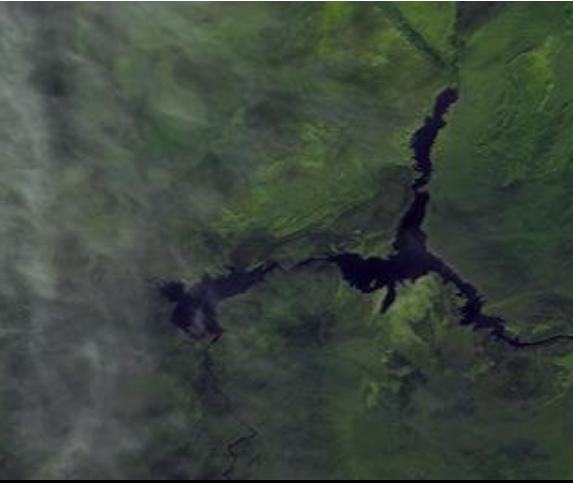} &
    \includegraphics[width=.32\linewidth,height=5cm]{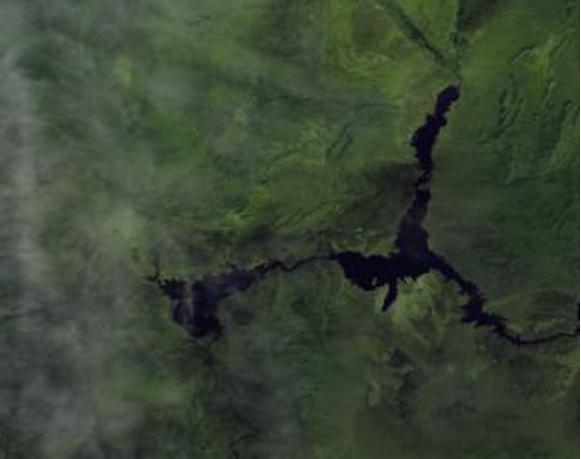}\\
    \end{tabularx}
    \end{center}
    \caption{ Left: Proba-V images. Center: Proba-V adapted with the $G_{\PV\to \LU}$ transformation ($A$ in Fig.\ref{fig:methodology}). Right: Landsat-8 upscaled image (LU in Fig.\ref{fig:methodology}).  Top: 1 day difference acquisitions from Nueva Esparta island in the Caribbean sea. Bottom: four minutes of difference acquisitions from Lake Made in North America.
    }
    \label{fig:textures}
\end{figure*}

\begin{figure}[t]
    \begin{tabularx}{\linewidth}{p{32mm}p{17mm}p{30mm}}
    \centering Proba-V (PV) & PV $\to$ L8 & L8-Upscaled (LU)\\
    \end{tabularx}
    \centering
    \includegraphics[width=.95\linewidth]{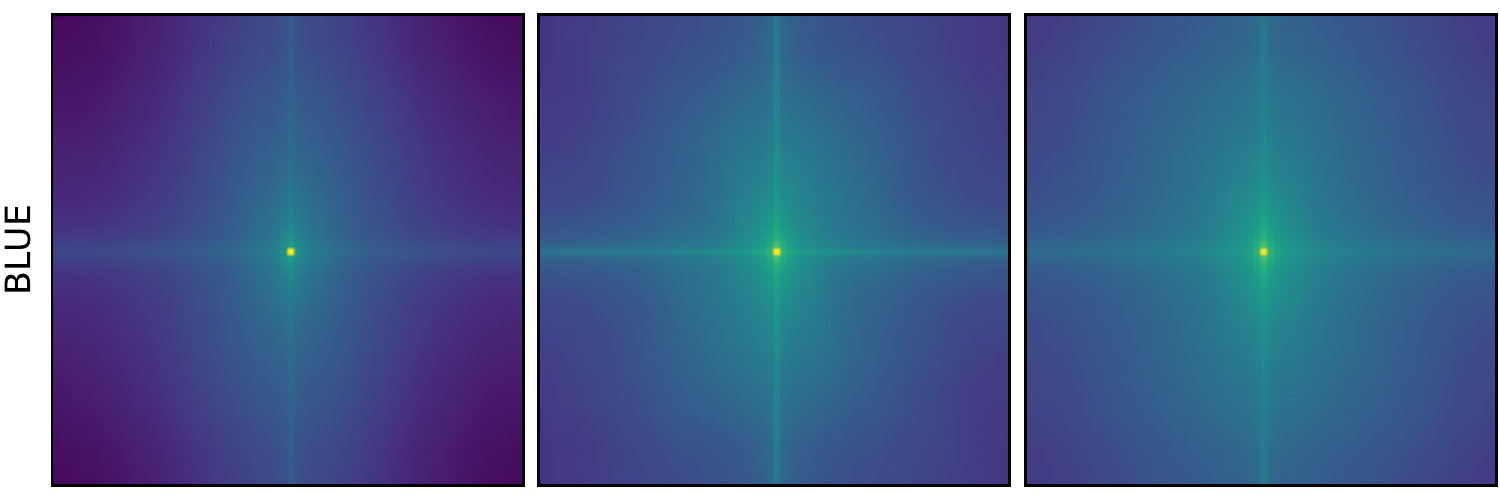}\\
    \includegraphics[width=.95\linewidth]{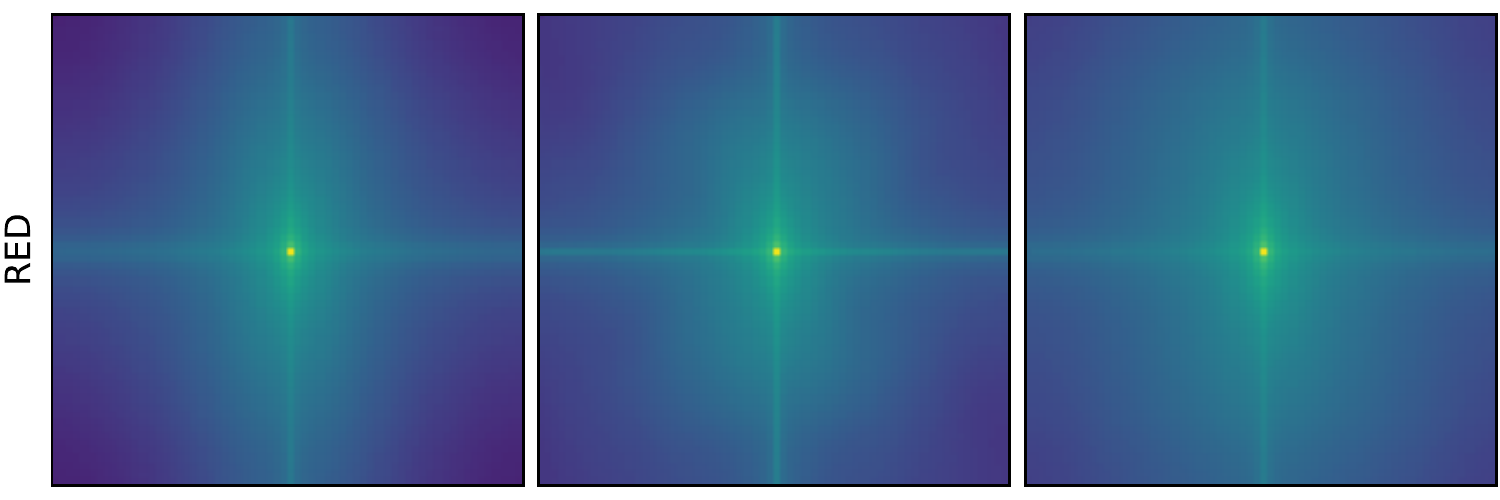}\\
    \includegraphics[width=.95\linewidth]{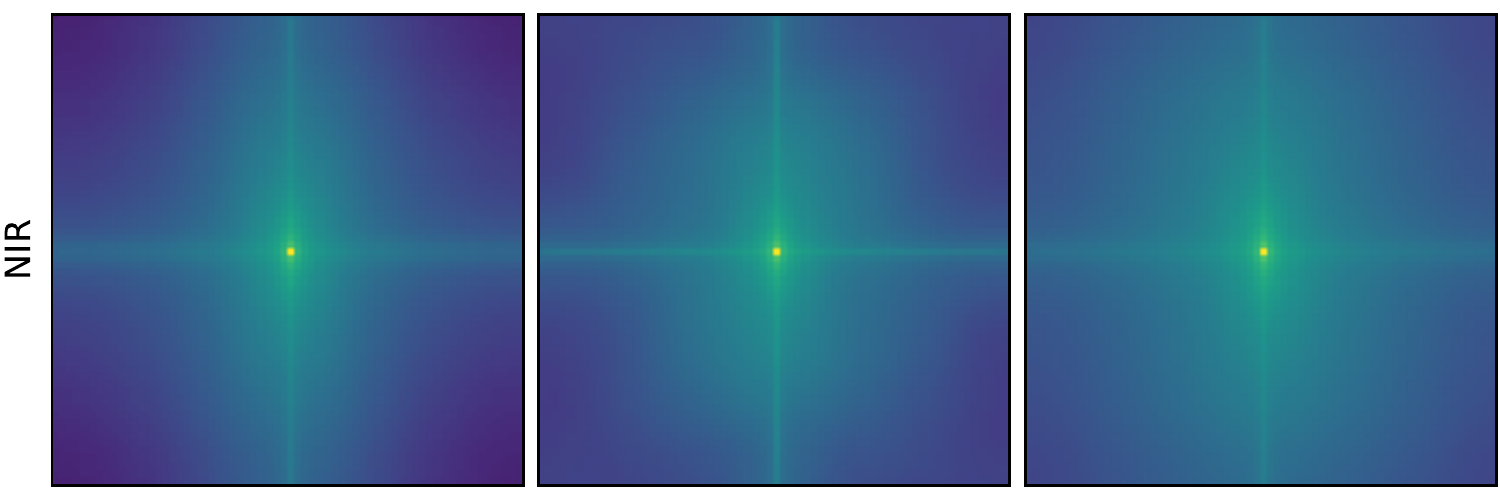}\\
    \includegraphics[width=.95\linewidth]{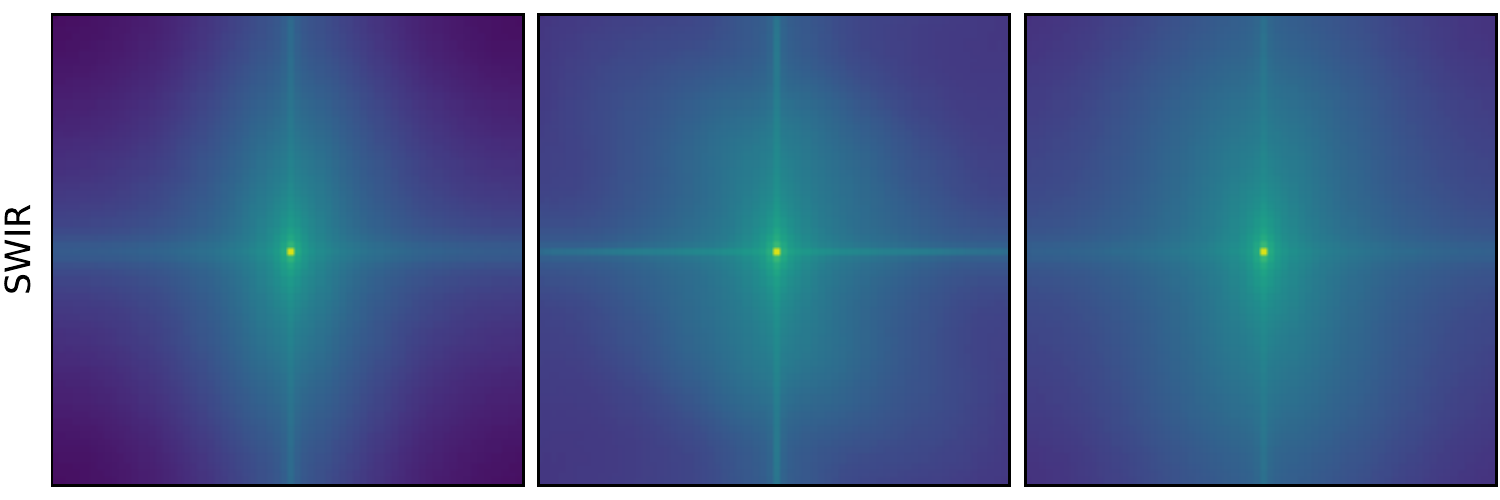}
    \caption{2D Fourier transform in dB for each of the four spectral channels averaged across all 64$\times$ 64 images patches in the \textit{38-Clouds Proba-V pseudo-simultaneous dataset}. Left: Proba-V images. Center: Proba-V images adapted using the proposed domain adaptation method. Right: pseudo-simultaneous Landsat-8 upscaled images.}
    \label{fig:fft}
\end{figure}

Visual examples of the trained DA network are shown in Fig.~\ref{fig:textures}. We show in the first column the Proba-V image, in the second one the adapted Proba-V image using our DA method, and in the third column the pseudo-simultaneous Landsat-8 upscaled image (LU). In the first row we can see that the location of clouds in the Proba-V image are preserved after the transformation while saturated blue values are removed. This provides a cloud appearance (and radiance) more similar to the pseudo-simultaneous LU images (third column). In the second row, we can see slightly sharper edges in the DA transformed image compared to the original Proba-V image. This is because the Landsat-8 upscaled images have components of higher spatial frequency than Proba-V. This was also point out in the pair of images at the bottom in Fig.~\ref{fig:pseudosimultaneouspvlandsat}. In order to test this hypothesis, 64$\times$64 pixels patches were extracted from the \textit{38-Clouds Proba-V pseudo-simultaneous dataset}. For each patch we computed the 2D Fast Fourier Transform for each of the four spectral bands. Finally, the amplitude of the signal at each frequency is converted to decibels (dB) and averaged across all patches (Fig.~\ref{fig:fft}). As pointed out before, Proba-V images have less high frequency components, whereas the average frequency amplitudes for the adapted images are more similar to the Landsat-8 upscaled ones. This highlights the spatio-spectral nature of the proposed method: it does not only learn spectral changes between bands (colors) but also spatial relations.

\begin{figure}[t]
    \centering
    \includegraphics[width=.75\linewidth]{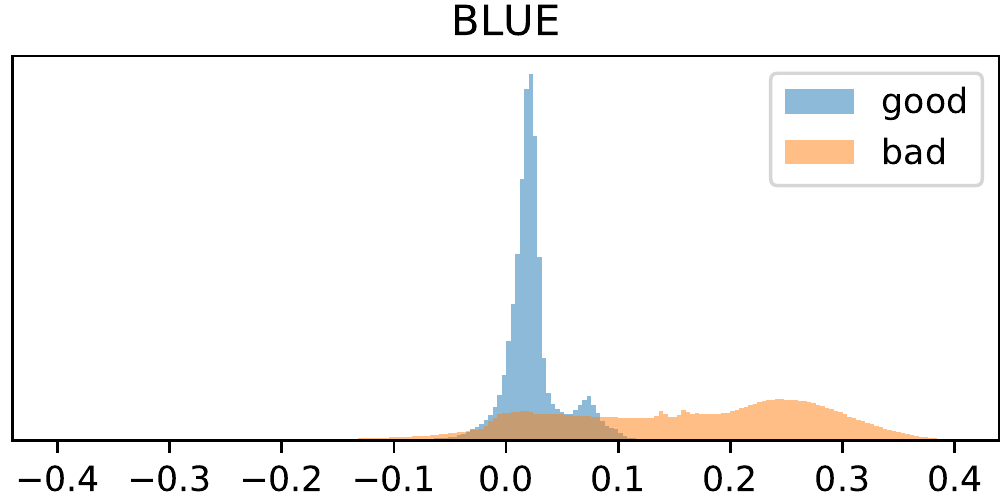}\\
    \includegraphics[width=.75\linewidth]{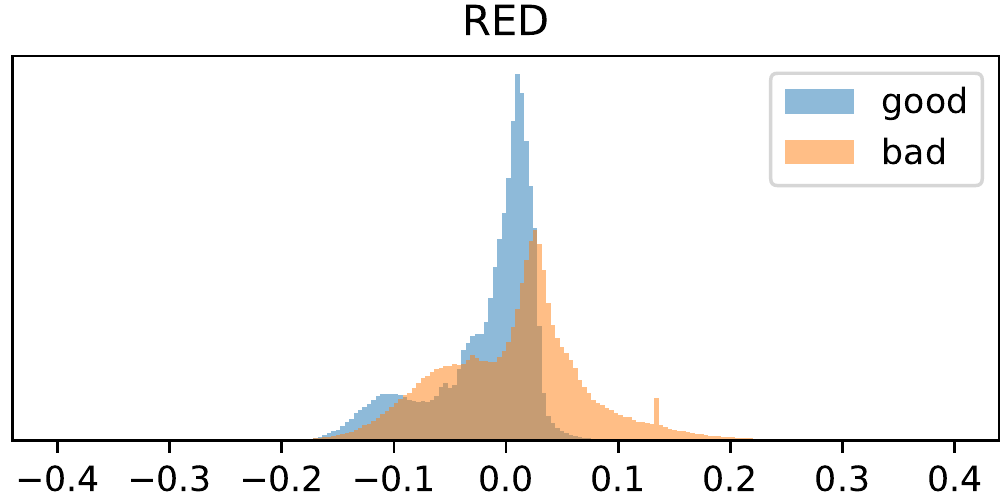}\\
    \includegraphics[width=.75\linewidth]{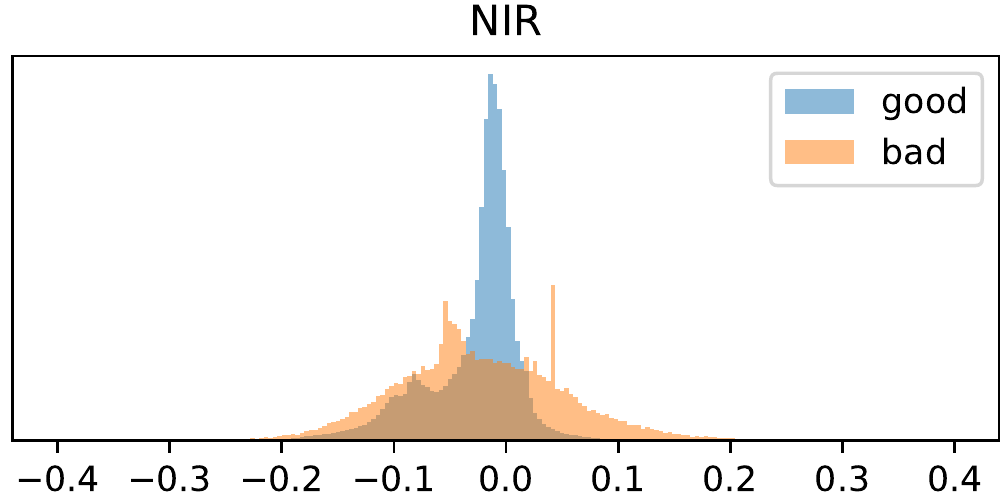}\\
    \includegraphics[width=.75\linewidth]{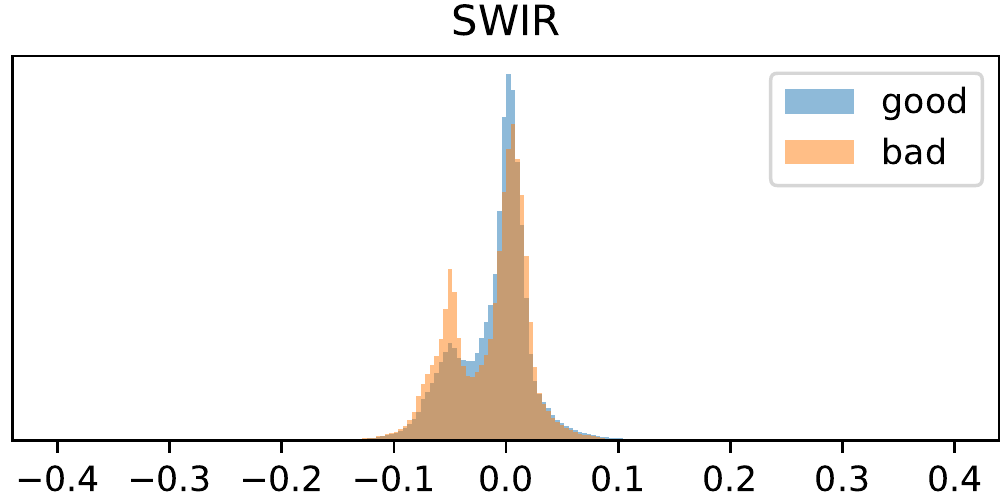}
    \caption{Differences in TOA for Proba-V images before and after applying the proposed DA transformation ($X_{\PV}-G_{\PV\to \LU}(X_{\PV})$)}
    \label{fig:difftoa}
\end{figure}

Finally, Fig.~\ref{fig:difftoa} shows the difference in TOA reflectance between the original Proba-V images and the adapted ones ($X_{\PV}-G_{\PV\to \LU}(X_{\PV})$) for all the pixels in the {\it 38-Clouds Proba-V pseudo-simultaneous dataset} and for each of the 4 Proba-V bands. In this case, we have stratified the pixels using the per pixel radiometric quality flag available in the status map (SM) of Proba-V products (see pag. 67 of Proba-V User Manual~\cite{Wolters18}). This quality indicator is a binary mask for each of the 4 Proba-V channels; pixels are flagged as \emph{bad quality} for different reasons, including detector saturation~\cite{Sterckx13}. In the Proba-V images of the {\it 38-Clouds Proba-V pseudo-simultaneous dataset}, approximately 30\% of pixels in the blue band have a reported bad quality, in contrast to the 5\% for the red band and 0.5\% for the NIR and SWIR. One can see that differences in TOA reflectance for saturated pixels is higher whereas good quality pixels change much less.




\subsection{Domain Adaptation for Cloud Detection} \label{sec:da4clouds}

In order to evaluate the DA methodology for the cloud detection application, we trained a FCNN in the \textit{Biome} dataset. In order to account for the uncertainty at the weights initialization and ordering of the batches, we trained 10 copies of the network using different random seed initializations. This procedure is also followed in \cite{mateo-garcia_transferring_2020}. Table~\ref{tab:results_source} shows the cloud detection accuracy on the source domain by using the \textit{SPARCS} and \textit{38-Clouds} datasets. We see that overall the accuracy is relatively high and networks are not much affected by the weights initialization.  

\begin{table}[ht]
    \centering
        \caption{Accuracy for test images in the source Landsat-8 Upscaled domain. Results averaged over 10 U-Net networks trained with different random initializations.}
    \label{tab:results_source}
\begin{tabularx}{.5\textwidth}{p{30mm}p{8mm}p{8mm}p{8mm}p{8mm}}
\toprule
Dataset &   min &   max &  mean &  std \\
\midrule
SPARCS & 91.67 & 92.46 & 92.12 & 0.20\\
38-Clouds & 90.32 &	91.92 & 91.31 & 0.47 \\
\bottomrule
\end{tabularx}
\end{table}

\begin{figure}[t]
    \centering
    \includegraphics[width=.49\linewidth,height=5cm]{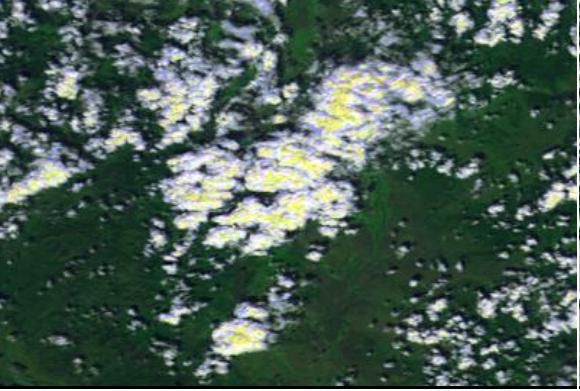}
    \includegraphics[width=.49\linewidth,height=5cm]{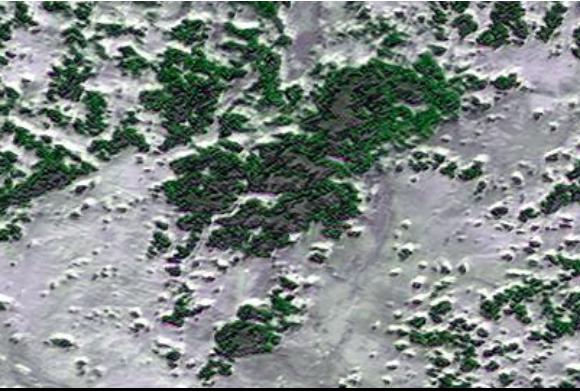}
    \caption{Example of color inversion when neither the segmentation loss nor the identity loss are included. Left: Proba-V image. Right: adapted image with the generator $G_{\PV\to \LU}$ trained with $\lambda_{seg}=0$, $\lambda_{id}=0$.}
    \label{fig:inversion}
\end{figure}

Table~\ref{tab:results} shows the results in the target domain (Proba-V) using the \textit{PV24} dataset with the trained DA transformation $G_{\PV\to \LU}$ (called \emph{full DA} in the table) and without it (called \emph{no DA}). We also include the results of the ablation study, where we have set some of the weights of the generator losses to zero and results using histogram matching~\cite{digitalimproc} for domain adaptation as in \cite{tasar_colormapgan_2020}. In addition, results are compared with the FCNN trained in original Proba-V images and ground truths (PV-trained), which serves as an upper bound reference, and with the operational Proba-V cloud detection algorithm (v101)~\cite{tote_evaluation_2018}. First of all, we see that the proposed DA method increases the mean overall accuracy and reduce the standard deviation of the metrics compared with direct transfer learning ({no DA}) or with adjusting the reflectance of each band with histogram matching. Secondly, we see minor reductions in accuracy when some losses are not used, in particular we see that the the segmentation loss and the identity loss have a strong influence in the result. We also see that when the segmentation and the identity loss are set to zero ($\lambda_{seg}=0$, $\lambda_{id}=0$) the cloud detection accuracy decreases abruptly. This is because without those losses generators are not constrained to maintain original radiance values and colors. A quick look at the generated images displayed in Fig.~\ref{fig:inversion} shows that the generators without these two loss terms could invert typical clear and cloud spectra.  

\begin{table}[ht]
    \centering
        \caption{Accuracy of different DA approaches for cloud detection over the \textit{PV24} dataset. Results averaged over 10 FCNN networks trained with different random initialization.}
    \label{tab:results}
\begin{tabularx}{.48\textwidth}{m{39mm}m{7mm}m{7mm}m{7mm}m{7mm}}
\toprule
{} &   min &   max &  mean &  std \\
\midrule
PV-trained~\cite{mateo-garcia_transferring_2020} & 94.83 & 95.15 & 94.98 & 0.09 \\
\midrule
full DA                  & \textbf{91.87} & \textbf{93.10} & \textbf{92.42} & 0.37  \\
$\lambda_{id}=0$ & \textbf{91.87} & 93.02 & 92.34 & 0.36 \\
$\lambda_{seg}=0$ & 91.15  & 91.86 & 91.47 & \textbf{0.24} \\
$\lambda_{cyc}=0$, $\lambda_{seg}=0$ & 90.35 & 91.26 & 90.79 & 0.32 \\
$\lambda_{seg}=0$, $\lambda_{id}=0$ & 33.99 & 36.10 & 35.20 & 0.63 \\
Histogram Matching~\cite{digitalimproc} & 89.09 & 91.31 & 90.18 & 0.72 \\
\midrule
no DA & 89.05 & 91.88 & 90.41 & 0.82 \\
\midrule
Proba-V operational v101~\cite{tote_evaluation_2018} & - & - & 82.97 & - \\
\bottomrule
\end{tabularx}
\end{table}

\begin{figure}[t]
    \centering
    \includegraphics[width=.85\linewidth]{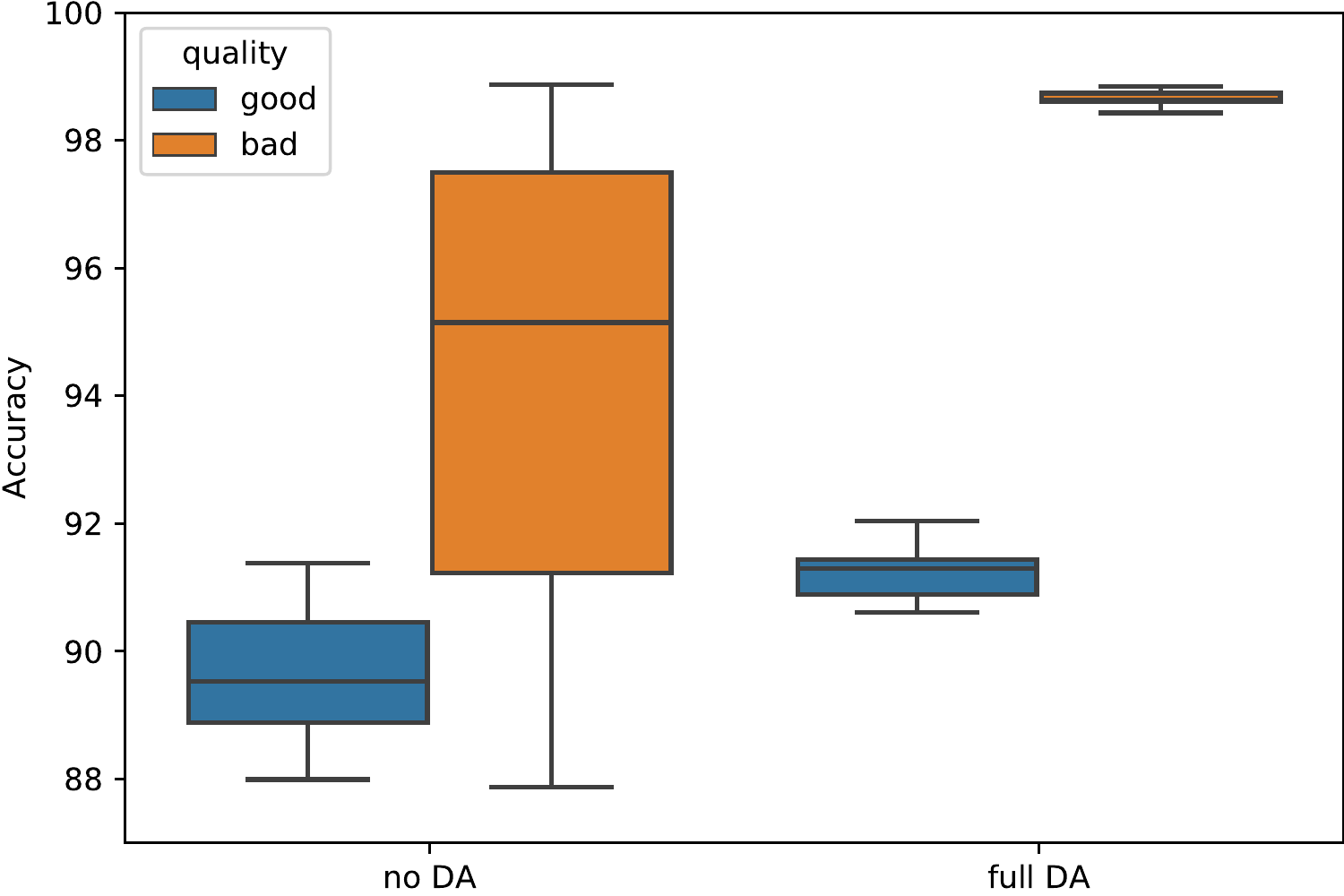}
    \caption{Cloud detection accuracy in the {\it PV-24} dataset of the 10 cloud detection U-Net models with different weight initialization with and without the proposed DA transformation. Pixels stratified according to the quality indicator available in the SM flag of Proba-V images.}
    \label{fig:boxplotacc}
\end{figure}

Figure~\ref{fig:boxplotacc} shows the cloud detection accuracy with the proposed DA transformation (full DA) and without DA (no DA). In this case, results are stratified using the quality flag available in the SM band of Proba-V. In the {\it PV-24} dataset, 16.13\% of pixels have at least one value in a band flagged as having bad quality. Within bad quality pixels, 96.8\% are cloudy pixels. We see that, on the one hand, if the DA transformation is not used, the accuracy of the networks wildly varies specially for bad quality pixels. These differences in accuracy of models trained on the same data indicate that the networks are extrapolating in those regions. On the other hand, when the DA transformation is used, all the networks identify correctly most of the bad quality pixels.

\begin{figure*}[t]
    \centering
    \begin{tabularx}{\linewidth}{p{36mm}p{34mm}p{28mm}p{30mm}p{30mm}}
    L8-Upscaled (LU) & Proba-V (PV) & PV $\to$ L8 & Clouds with DA & Clouds without DA\\
    \end{tabularx}
    \includegraphics[width=.19\linewidth]{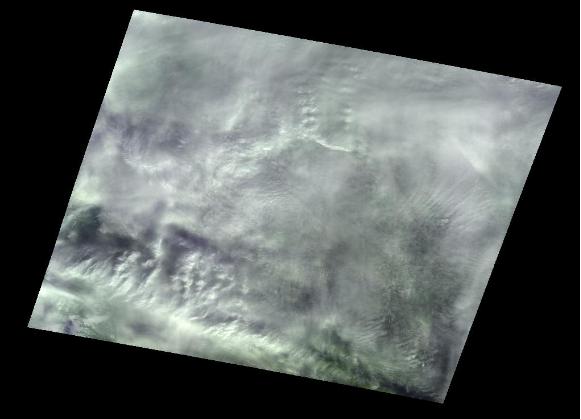}
    \includegraphics[width=.19\linewidth]{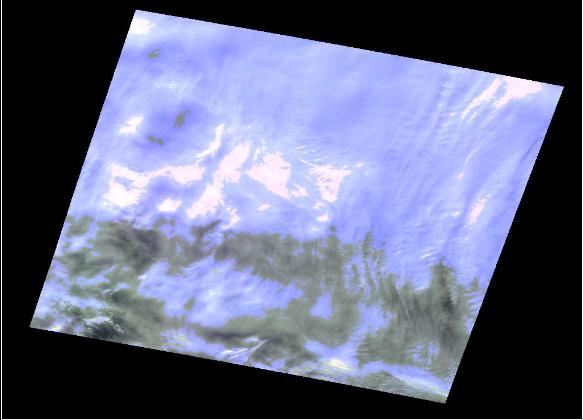}
    \includegraphics[width=.19\linewidth]{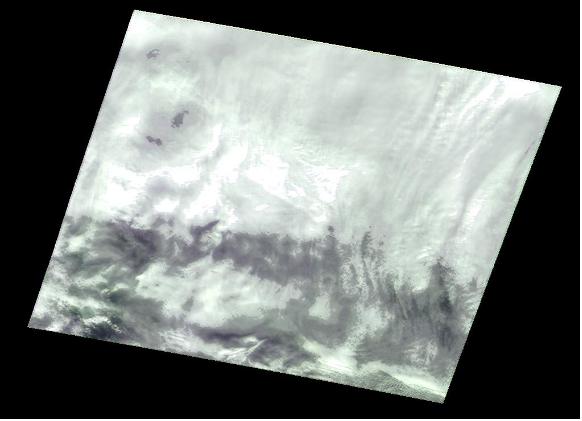}
    \includegraphics[width=.19\linewidth]{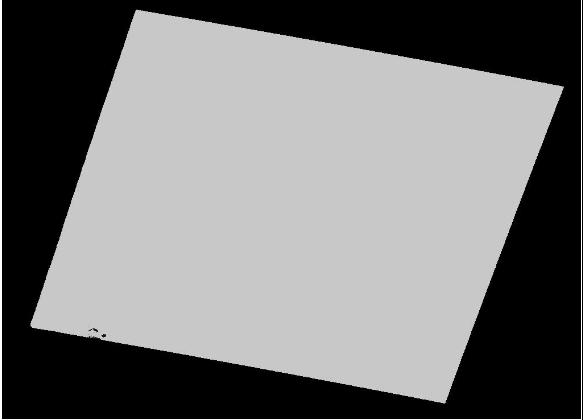}
    \includegraphics[width=.19\linewidth]{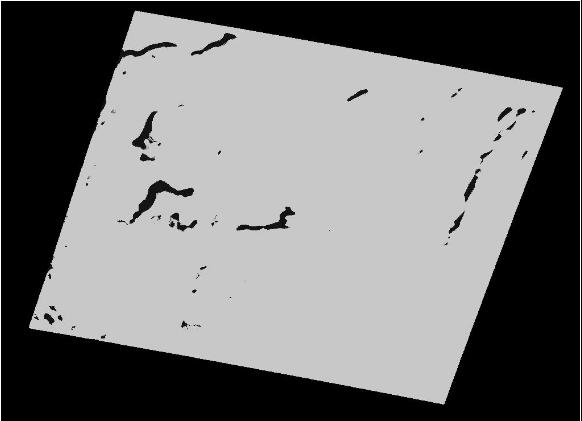}\\
    \includegraphics[width=.19\linewidth,height=3cm]{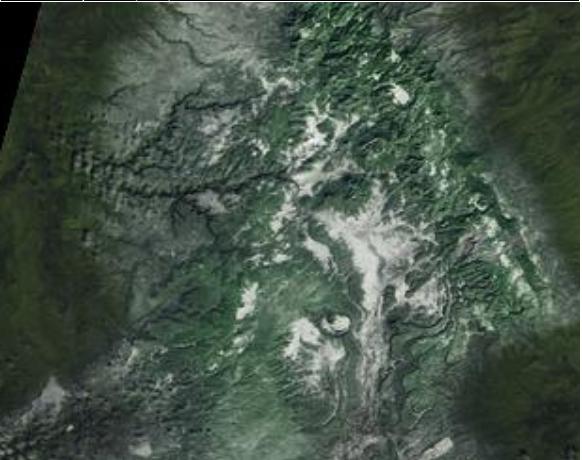}
    \includegraphics[width=.19\linewidth,height=3cm]{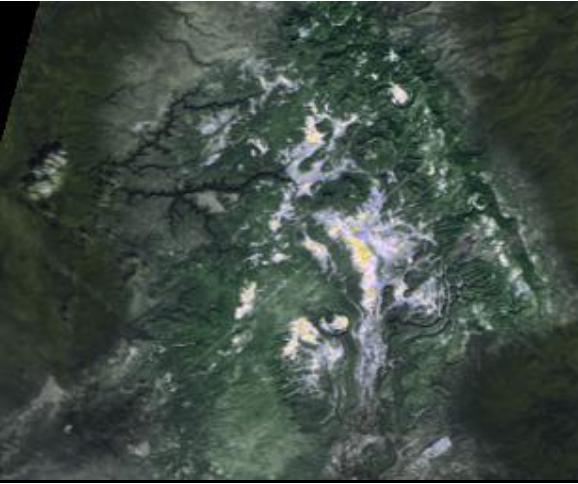}
    \includegraphics[width=.19\linewidth,height=3cm]{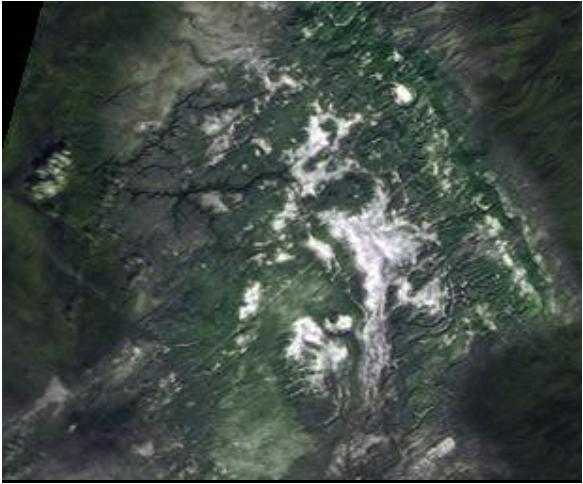}
    \includegraphics[width=.19\linewidth,height=3cm]{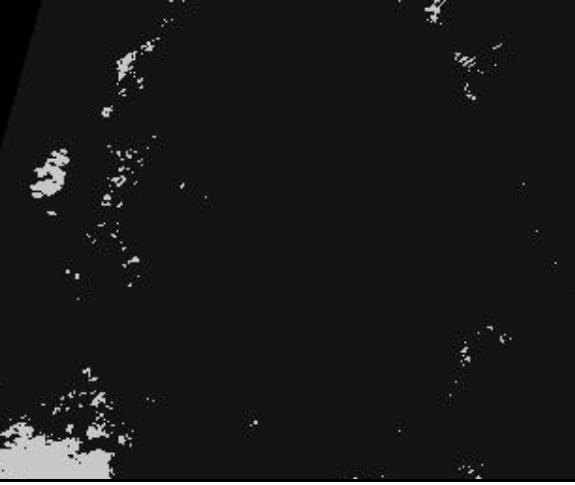}
    \includegraphics[width=.19\linewidth,height=3cm]{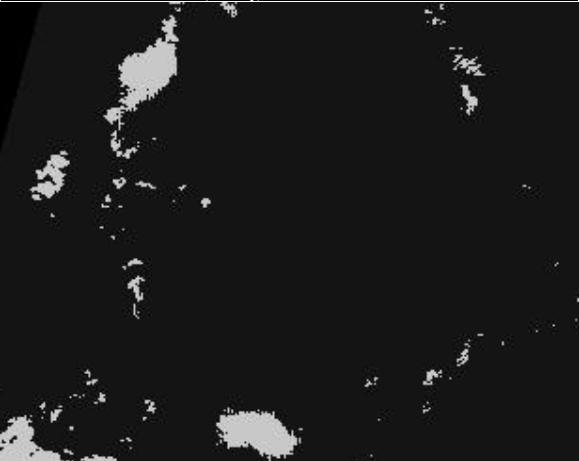}\\
     \includegraphics[width=.19\linewidth,height=3cm]{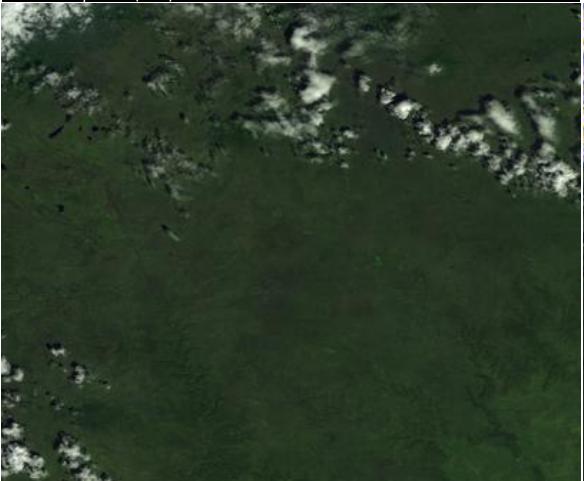}
    \includegraphics[width=.19\linewidth,height=3cm]{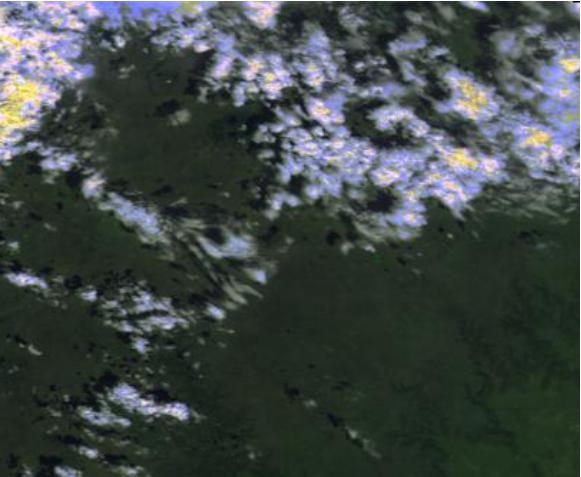}
    \includegraphics[width=.19\linewidth,height=3cm]{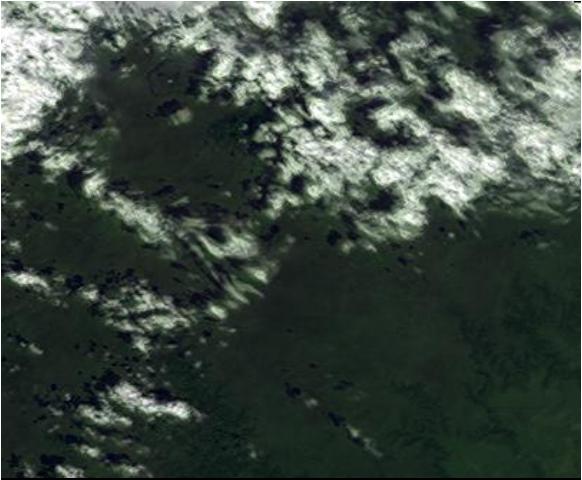}
    \includegraphics[width=.19\linewidth,height=3cm]{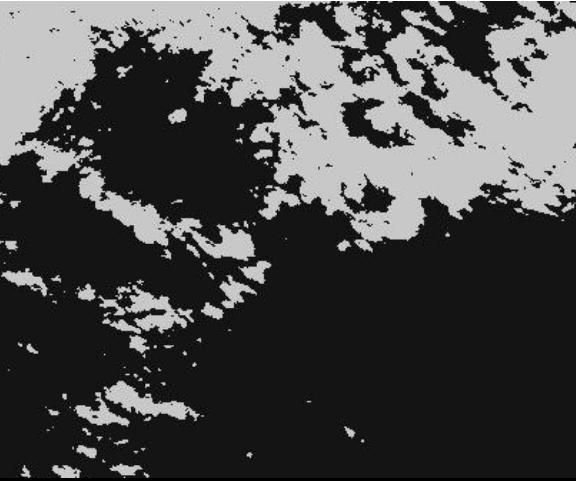}
    \includegraphics[width=.19\linewidth,height=3cm]{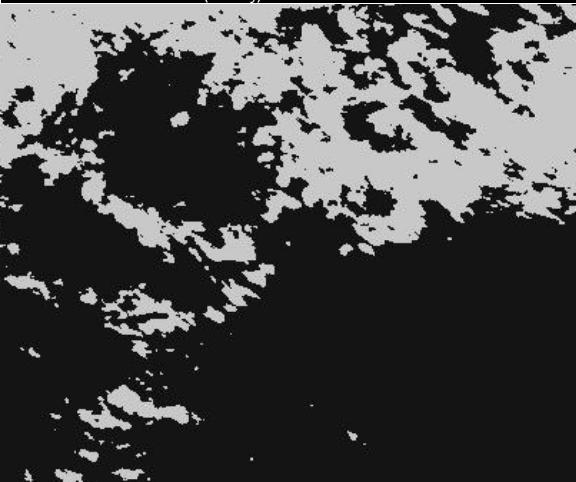}\\
    \includegraphics[width=.19\linewidth,height=3cm]{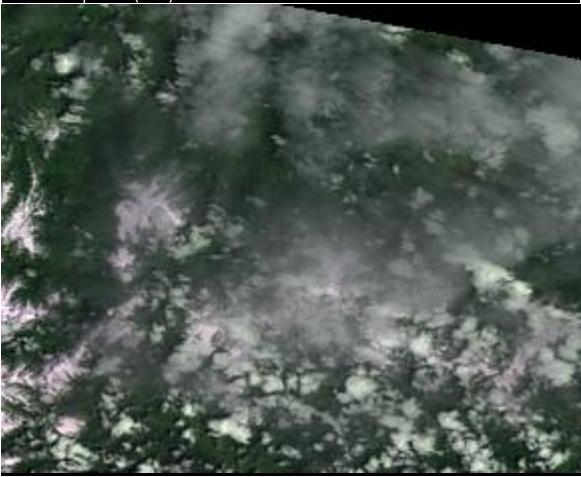}
    \includegraphics[width=.19\linewidth,height=3cm]{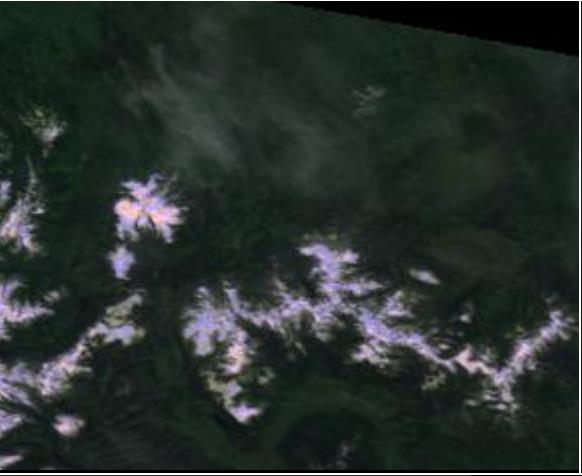}
    \includegraphics[width=.19\linewidth,height=3cm]{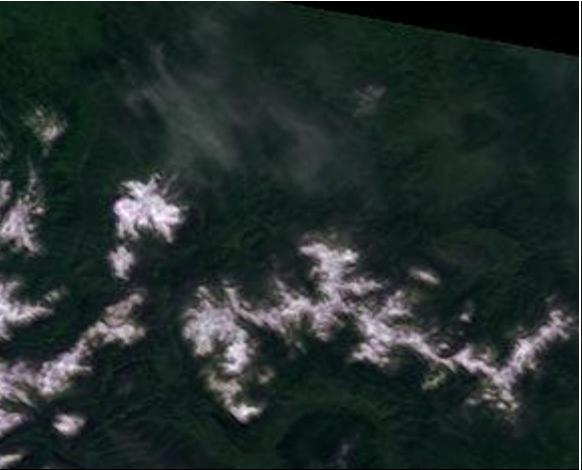}
    \includegraphics[width=.19\linewidth,height=3cm]{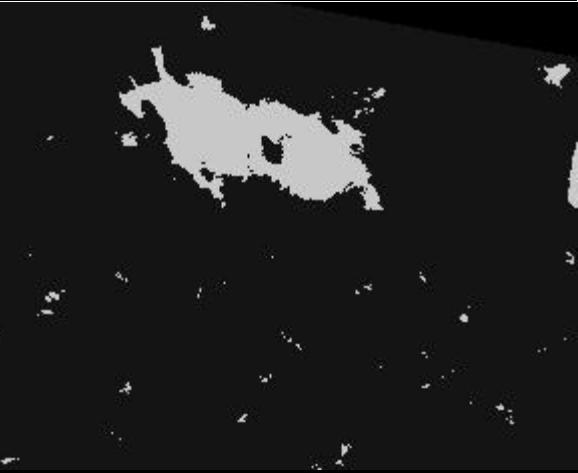}
    \includegraphics[width=.19\linewidth,height=3cm]{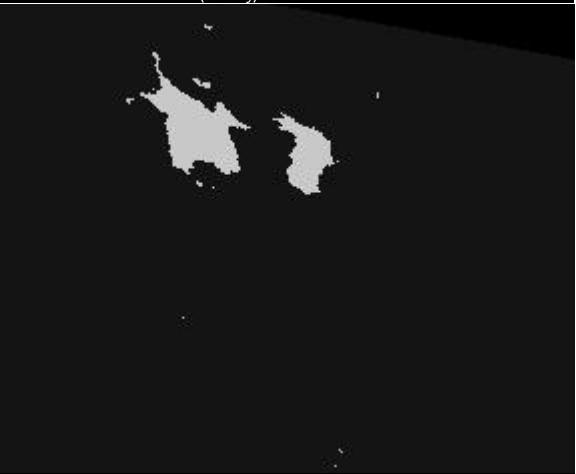}\\
    \caption{From left to right: Landsat-8 Upscaled image, Proba-V image, Proba-V as Landsat-8 upscaled, Clouds from Proba-V as Landsat-8 upscaled, Clouds without domain adaptation.}
    \label{fig:samplesclouds}
\end{figure*}

Finally, Fig.~\ref{fig:samplesclouds} shows some cherry-picked examples of cloud detection with the proposed methodology. On each row we show: the pseudo-simultaneous Landsat-8 Upscaled image, the original Proba-V image, the Proba-V image after the domain adaptation $G_{\PV\to \LU}$, the cloud mask using as input the DA image, and the cloud mask obtained without the domain adaptation. First row shows a completely cloudy image with several blue saturated pixels in the original Proba-V image. We see that the DA image removes those saturated values and helps the cloud detection model to correctly predict all pixels. We see that, if no DA transformation is employed, the saturated values in Proba-V hinder the performance of the model with some cloud miss-classifications. The second row shows an acquisition over the Canyon de Chelly, in North America. We see again that saturated values in the blue band disappear after the DA transformation; in this case, this help to reduce the false positives in the bottom and upper left part of the image. In the third row, we see an easier case where cloud masks, with and without DA, are both accurate. Finally, in the fourth row, a very challenging example of thin clouds over snowy mountains is shown. In this case, the DA method captures better the thin cloud in the top of the image; however, it produces some false positives in mixed pixels where snow is melting. For results of all methods shown in Table~\ref{tab:results} over all the images in the \textit{38-Cloud Proba-V pseudo-simultaneous dataset}, we refer the reader to the web application in~\url{https://isp.uv.es/projects/cloudsat/pvl8dagans}.

\section{Discussion and conclusions} \label{sec:conclusions}

The main motivation for this study has been to propose a domain adaptation algorithm to improve transfer learning from Landsat-8 to Proba-V for cloud detection. 
However, the obtained transformation is application-independent since its aim is to reduce spatial and spectral differences between the two sensors image datasets. 
It is worth noting that the main objective of Proba-V mission was to ensure continuity and fill the gap between SPOT Vegetation (VGT) and the Sentinel-3 missions~\cite{ToteIGARSS18}.
These multi-mission approaches are necessary to provide long time-series of surface vegetation products with complete Earth coverage, and require a high radiometric consistency across sensor datasets. For instance, differences between Proba-V and VGT2 spectral responses were of the same order as between VGT1 and VGT2~\cite{sterckx_orbit_2016}. Also, the ESA Sentinel-3 synergy vegetation products replicate the characteristics of the 1-km SPOT VGT products using spectral remapping and co-location techniques. 
In this context, the proposed domain adaptation model is useful for any general purpose cross-sensor application. 
In addition, it directly learns from the satellite images and is based on sound statistical methods.
Therefore, the transformation is suitable for general use and could be applied to any remote sensing problem in which Proba-V could take advantage of Landsat-8 data.

Obtained results have shown that the proposed domain adaptation transformation, in addition to reduce the difference between the TOA reflectance distributions, also unintentionally fixes some radiometry artifacts. Looking at the resulting distributions one can see that the characteristic saturation in the Blue and Red channels of Proba-V disappears in the adapted images (see Fig.~\ref{fig:histograms}).
Moreover, although the developed model does not distinguish explicitly between \emph{good} and \emph{bad} pixels of the Proba-V products quality flag, results show that good quality pixels are much less changed than bad quality pixels (see Fig.~\ref{fig:difftoa}). On the one hand, this result agrees with~\cite{sterckx_radiometric_2019}, where good quality pixels are similar between Landsat-8 and Proba-V, which implies that their TOA radiance calibration is quite good. On the other hand, the proposed adaptation method only changes those \emph{good} pixels within the range of the radiometric error reported in~\cite{sterckx_radiometric_2019} or in~\cite{Sterckx13} (see Fig.~\ref{fig:difftoa}). 
However, Proba-V products present a significant number of bad quality pixels: between 20\% and 30\% in the blue channel and around 5\% in the red one. This can eventually have an important impact on derived products, since usually we are expected to provide results in the whole image. For instance, removing or ignoring those \emph{bad} pixels is not feasible for methods using the spatial context, such as CNNs, since the output for a pixel depends on the surrounding pixels. Therefore, the DA method improves the TOA reflectance image resemblance across sensors and, in this particular case, significantly increases the number of pixels that can be further processed: i.e. corrected bad quality pixels (see Fig.~\ref{fig:boxplotacc}).

In addition, the proposed adversarial domain adaptation model has been modified to specifically improve the performance of a transfer learning cloud detection problem. In particular, the cost function used to train the DA network has been modified by including a dedicated term forcing similar cloud detection results across domains. Results show that, when the proposed transformation is applied, cloud detection models trained using only Landsat-8 data increase cloud detection accuracy in Proba-V.
It is worth noting that results without the application dependent term ($\lambda_{seg}=0$) are good enough. However, it is important to include either $\lambda_{seg}>0$ or $\lambda_{id}>0$ in order to constrain the method and to avoid artifacts in the adapted images.

The proposed adaptation framework can be extended in two ambitious directions.
On the one hand, it would be possible to learn a domain adaptation transformation directly from Landsat-8 to Proba-V, without previously applying the upscaling transformation, which converted the Landsat-8 images in order to have similar spatio-spectral properties than Proba-V (number of bands and spatial resolution). 
However, this approach would imply to solve the more challenging super-resolution problem when transforming Proba-V to Landsat-8 in our cyclic GAN adaptation framework. 
On the other hand, an interesting option to explore would be to apply the transformation from top of atmosphere to surface reflectance data. In this case, the obtained transformation would be equivalent to learn an atmospheric correction transformation relying on the image data only. 
In addition, as mentioned before, the proposed framework can be applied to any other pair of similar sensors such as Proba-V and Sentinel 2 and 3.




Summarizing, in this paper, a Cycle-GAN architecture has been proposed to train a domain adaptation method between Proba-V and upscaled Landsat images. The proposal includes two generators, two discriminators and four different penalties. The GAN generator is used to modify the Proba-V images to better resemble the upscaled Landsat images that have also been used to train a cloud detection algorithm. Results on original Proba-V images demonstrate that when using the proposed model for the adaptation a higher cloud detection accuracy is achieved. 

\appendices

\section{}
\label{sec:gandetails}

This appendix presents the details about the network architectures and the training procedure of the generators and discriminators of the proposed generative adversarial adaptation model. It also has the details of the networks and training configuration of the cloud detection models. The implementation is available at~\url{https://github.com/IPL-UV/pvl8dagans}.

\begin{figure}[ht]
    \centering
    \includegraphics[width=.8\linewidth]{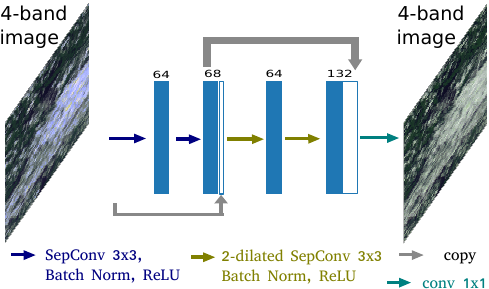}\\
    \includegraphics[width=.8\linewidth]{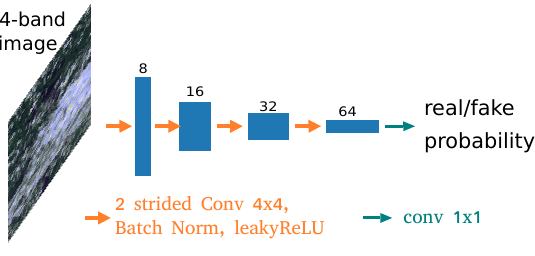}
    \caption{Generator (top) and discriminator (bottom) architectures. Implementation details available at \url{https://github.com/IPL-UV/pvl8dagans}}
    \label{fig:discgen}
\end{figure}

We use the same network architecture for all generators $G$ (fig.~\ref{fig:discgen} top). In particular, $G$ is a 5-layer fully convolutional neural network . It consists of:
\begin{itemize}
    \item 2 layers: convolution with 64 separable filters of size 3$\times$3, reLU activation, and batch normalization.
    \item 2 layers: convolution with 64 separable filters of size 3$\times$3 with a dilation rate equal to 2, reLU activation, and batch normalization.
    \item 1 layer: 1$\times$1 convolution with 4 channels output.
\end{itemize} 
We used residual connections between blocks and before the final layer.

As in the case of the generators, both discriminators, $D$, have the same architecture: a 5-layer convolutional neural network adapted from~\cite{isola2017image} (fig.~\ref{fig:discgen} bottom). It consists of:
\begin{itemize}
    \item 4 layers: 4$\times$4 convolution, leakyReLU activation and batch normalization. The number of filters starts in 8 for the first convolution an grows by a factor two in every layer. The convolutions are applied with a stride of 2, thus reducing by this factor the spatial size of the input. 
    \item 1 layer: 1$\times$1 convolution with 1 output channel and a sigmoid activation.
\end{itemize} 
The output of the discriminators can be interpreted as the probability of an image to be \emph{real}. Hence, the discriminator is trained to provide close-to-zero values for images generated by $G$ and close-to-one values for real satellite images.


The proposed networks ($G_{\PV\to \LU}$, $G_{\LU\to \PV}$, $D_{\PV}$ and $D_{\LU}$) were trained simultaneously using 64$\times$ 64 patches with stochastic gradient descent on their respective losses with a batch size of 48. Networks were trained for 25 epochs in the {\it Proba-V pseudo-simultaneous dataset}, which corresponds to 14,574 steps where the weights are updated. In order to ensure convergence in the GAN training procedure, we regularized the discriminator using 0 centered gradient penalty on the real images~\cite{Mescheder2018ICML} with a weight of 10. We used the the Adam~\cite{KingmaB14} optimizer with a learning rate of $10^{-4}$ to update the weights of the networks at each step. Additionally, we apply data augmentations in form of 90 degree rotations and horizontal and vertical flips.

\begin{figure}[t]
    \centering
    \includegraphics[width=\linewidth]{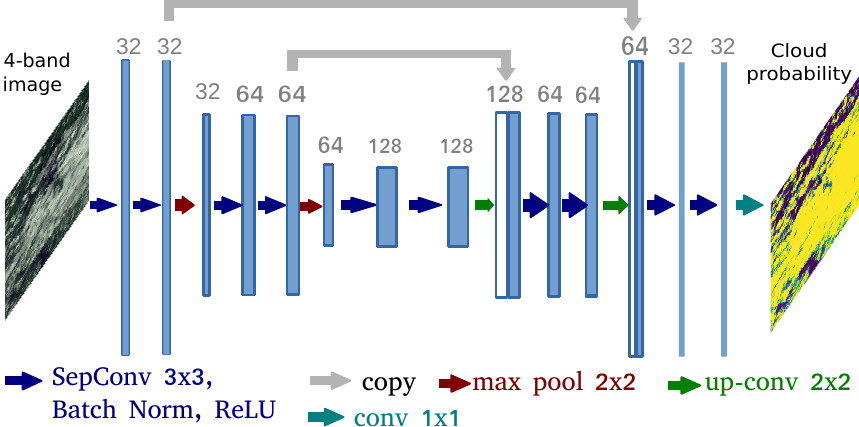}
    \caption{Simplified U-Net architecture used for the cloud detection model. Implementation details available at \url{https://github.com/IPL-UV/pvl8dagans}}
    \label{fig:unet_arch}
\end{figure}

For the cloud detection model $f_{LU}$, we used the same simplified U-Net architecture as in the work~\cite{mateo-garcia_transferring_2020}. Figure~\ref{fig:unet_arch} shows the configuration of layers; we used only two subsampling steps and separable convolutions~\cite{chollet_xception_2017} to reduce the number of trainable parameters and floating points operations (96k parameters and 2.18M FLOPS). The cloud detection networks are trained for 250k steps using batches of 64 overlapping patches of 32$\times$ 32 pixels from the Biome dataset (upscaled to 333m as described in sec.~\ref{sec:datasets}). We used a learning rate of $10^{-4}$ and the Adam~\cite{KingmaB14} optimizer.



\bibliographystyle{IEEEtran}
\bibliography{biblio}

\begin{thebibliography}{10}
\providecommand{\url}[1]{#1}
\csname url@samestyle\endcsname
\providecommand{\newblock}{\relax}
\providecommand{\bibinfo}[2]{#2}
\providecommand{\BIBentrySTDinterwordspacing}{\spaceskip=0pt\relax}
\providecommand{\BIBentryALTinterwordstretchfactor}{4}
\providecommand{\BIBentryALTinterwordspacing}{\spaceskip=\fontdimen2\font plus
\BIBentryALTinterwordstretchfactor\fontdimen3\font minus
  \fontdimen4\font\relax}
\providecommand{\BIBforeignlanguage}[2]{{%
\expandafter\ifx\csname l@#1\endcsname\relax
\typeout{** WARNING: IEEEtran.bst: No hyphenation pattern has been}%
\typeout{** loaded for the language `#1'. Using the pattern for}%
\typeout{** the default language instead.}%
\else
\language=\csname l@#1\endcsname
\fi
#2}}
\providecommand{\BIBdecl}{\relax}
\BIBdecl

\bibitem{ucsatellite}
``{UCS Satellite Database},''
  \url{https://www.ucsusa.org/resources/satellite-database}, accessed:
  2020-04-25.

\bibitem{DARSTuia2016}
D.~Tuia, C.~Persello, and L.~Bruzzone, ``Domain adaptation for the
  classification of remote sensing data: An overview of recent advances,''
  \emph{IEEE Geoscience and Remote Sensing Magazine}, vol.~4, pp. 41--57, 06
  2016.

\bibitem{torralba_unbiased_2011}
A.~Torralba and A.~A. Efros, ``Unbiased look at dataset bias,'' in \emph{{CVPR}
  2011}, Jun. 2011, pp. 1521--1528.

\bibitem{mateo-garcia_domain_2019}
G.~Mateo-García, V.~Laparra, and L.~Gómez-Chova, ``Domain {Adaptation} of
  {Landsat}-8 and {Proba}-{V} {Data} {Using} {Generative} {Adversarial}
  {Networks} for {Cloud} {Detection},'' in \emph{{IGARSS} 2019 - 2019 {IEEE}
  {International} {Geoscience} and {Remote} {Sensing} {Symposium}}, Jul. 2019,
  pp. 712--715, iSSN: 2153-6996.

\bibitem{mateo-garcia_transferring_2020}
G.~Mateo-García, V.~Laparra, D.~López-Puigdollers, and L.~Gómez-Chova,
  ``\BIBforeignlanguage{en}{Transferring deep learning models for cloud
  detection between {Landsat}-8 and {Proba}-{V}},''
  \emph{\BIBforeignlanguage{en}{ISPRS Journal of Photogrammetry and Remote
  Sensing}}, vol. 160, pp. 1--17, Feb. 2020.

\bibitem{Banerjee17JSTARS}
B.~{Banerjee} and S.~{Chaudhuri}, ``Hierarchical subspace learning based
  unsupervised domain adaptation for cross-domain classification of remote
  sensing images,'' \emph{IEEE Journal of Selected Topics in Applied Earth
  Observations and Remote Sensing}, vol.~10, no.~11, pp. 5099--5109, Nov 2017.

\bibitem{svendsen_joint_2019}
D.~H. {Svendsen}, L.~{Martino}, M.~{Campos-Taberner}, F.~J. {García-Haro}, and
  G.~{Camps-Valls}, ``Joint gaussian processes for biophysical parameter
  retrieval,'' \emph{IEEE Transactions on Geoscience and Remote Sensing},
  vol.~56, no.~3, pp. 1718--1727, March 2018.

\bibitem{wolanin_estimating_2020}
A.~Wolanin, G.~Mateo-García, G.~Camps-Valls, L.~Gómez-Chova, M.~Meroni,
  G.~Duveiller, Y.~Liangzhi, and L.~Guanter,
  ``\BIBforeignlanguage{en}{Estimating and understanding crop yields with
  explainable deep learning in the {Indian} {Wheat} {Belt}},''
  \emph{\BIBforeignlanguage{en}{Environmental Research Letters}}, vol.~15,
  no.~2, p. 024019, Feb. 2020.

\bibitem{Landsat8handbook}
{U.S. Geological Survey}, ``{Landsat 8 Data Users Handbook},'' USGS, Tech. Rep.
  LSDS-1574, April 2019,
  https://www.usgs.gov/media/files/landsat-8-data-users-handbook.

\bibitem{Dierckx14}
W.~Dierckx, S.~Sterckx, I.~Benhadj, S.~Livens, G.~Duhoux, T.~Van~Achteren,
  M.~Francois, K.~Mellab, and G.~Saint, ``{PROBA-V} mission for global
  vegetation monitoring: standard products and image quality,'' \emph{Int.
  Journal of Remote Sensing}, vol.~35, no.~7, pp. 2589--2614, 2014.

\bibitem{sterckx_radiometric_2019}
S.~Sterckx and E.~Wolters, ``\BIBforeignlanguage{en}{Radiometric
  {Top}-of-{Atmosphere} {Reflectance} {Consistency} {Assessment} for {Landsat}
  8/{OLI}, {Sentinel}-2/{MSI}, {PROBA}-{V}, and {DEIMOS}-1 over {Libya}-4 and
  {RadCalNet} {Calibration} {Sites}},'' \emph{\BIBforeignlanguage{en}{Remote
  Sensing}}, vol.~11, no.~19, p. 2253, Jan. 2019.

\bibitem{Sterckx13}
\BIBentryALTinterwordspacing
S.~Sterckx, W.~Dierckx, S.~Adriaensen, and S.~Livens, ``{PROBA-V Commissioning
  Report Annex 1-Radiometric Calibration Results},'' VITO, Tech. Rep. Technical
  Note, 11/2013, Nov 2013. [Online]. Available:
  \url{https://earth.esa.int/documents/700255/1929094/US-20+Annex1-RadiometricCalibartion-v1_1.pdf/389c059f-4808-4f92-9642-2cab5a4450cb}
\BIBentrySTDinterwordspacing

\bibitem{hoffman_cycada_2018}
J.~Hoffman, E.~Tzeng, T.~Park, J.-Y. Zhu, P.~Isola, K.~Saenko, A.~Efros, and
  T.~Darrell, ``\BIBforeignlanguage{en}{{CyCADA}: {Cycle}-{Consistent}
  {Adversarial} {Domain} {Adaptation}},'' in
  \emph{\BIBforeignlanguage{en}{{ICML} 2018}}, Jul. 2018, pp. 1989--1998.

\bibitem{CycleGAN2017}
J.-Y. Zhu, T.~Park, P.~Isola, and A.~A. Efros, ``Unpaired image-to-image
  translation using cycle-consistent adversarial networkss,'' in \emph{Computer
  Vision (ICCV), 2017 IEEE International Conference on}, 2017.

\bibitem{long_fully_2015}
J.~Long, E.~Shelhamer, and T.~Darrell, ``Fully convolutional networks for
  semantic segmentation,'' in \emph{2015 {IEEE} {Conference} on {Computer}
  {Vision} and {Pattern} {Recognition} ({CVPR})}, Jun. 2015, pp. 3431--3440.

\bibitem{ronneberger_u-net_2015}
O.~Ronneberger, P.~Fischer, and T.~Brox, ``\BIBforeignlanguage{en}{U-{Net}:
  {Convolutional} {Networks} for {Biomedical} {Image} {Segmentation}},'' in
  \emph{\BIBforeignlanguage{en}{Medical {Image} {Computing} and
  {Computer}-{Assisted} {Intervention} – {MICCAI} 2015}}, ser. LNCS.\hskip
  1em plus 0.5em minus 0.4em\relax Springer, Cham, Oct. 2015, pp. 234--241.

\bibitem{RSNet}
J.~H. Jeppesen, R.~H. Jacobsen, F.~Inceoglu, and T.~S. Toftegaard, ``A cloud
  detection algorithm for satellite imagery based on deep learning,''
  \emph{Remote Sensing of Environment}, vol. 229, pp. 247--259, Aug. 2019.

\bibitem{38-cloud-1}
S.~{Mohajerani} and P.~{Saeedi}, ``Cloud-net: An end-to-end cloud detection
  algorithm for landsat 8 imagery,'' in \emph{IGARSS 2019 - 2019 IEEE
  International Geoscience and Remote Sensing Symposium}, July 2019, pp.
  1029--1032.

\bibitem{38-cloud-2}
S.~Mohajerani, T.~A. Krammer, and P.~Saeedi, ``{"A Cloud Detection Algorithm
  for Remote Sensing Images Using Fully Convolutional Neural Networks"},'' in
  \emph{2018 IEEE 20th International Workshop on Multimedia Signal Processing
  (MMSP)}, Aug 2018, pp. 1--5.

\bibitem{li_deep_2019}
Z.~Li, H.~Shen, Q.~Cheng, Y.~Liu, S.~You, and Z.~He, ``Deep learning based
  cloud detection for medium and high resolution remote sensing images of
  different sensors,'' \emph{ISPRS Journal of Photogrammetry and Remote
  Sensing}, vol. 150, pp. 197--212, Apr. 2019.

\bibitem{yang_cdnet:_2019}
J.~Yang, J.~Guo, H.~Yue, Z.~Liu, H.~Hu, and K.~Li, ``{CDnet}: {CNN}-{Based}
  {Cloud} {Detection} for {Remote} {Sensing} {Imagery},'' \emph{IEEE
  Transactions on Geoscience and Remote Sensing}, pp. 1--17, 2019.

\bibitem{zhu_improvement_2015}
Z.~Zhu, S.~Wang, and C.~E. Woodcock, ``Improvement and expansion of the {Fmask}
  algorithm: cloud, cloud shadow, and snow detection for {Landsats} 4–7, 8,
  and {Sentinel} 2 images,'' \emph{Remote Sensing of Environment}, vol. 159,
  no. Supplement C, pp. 269--277, Mar 2015.

\bibitem{wieland_multi-sensor_2019}
M.~Wieland, Y.~Li, and S.~Martinis, ``Multi-sensor cloud and cloud shadow
  segmentation with a convolutional neural network,'' \emph{Remote Sensing of
  Environment}, vol. 230, p. 111203, Sep. 2019.

\bibitem{geological_survey_l8_2016}
\BIBentryALTinterwordspacing
U.~Geological~Survey, ``L8 {SPARCS} {Cloud} {Validation} {Masks},'' \emph{U.S.
  Geological Survey}, no. data release, 2016. [Online]. Available:
  \url{https://landsat.usgs.gov/sparcs}
\BIBentrySTDinterwordspacing

\bibitem{segal-rozenhaimer_cloud_2020}
M.~Segal-Rozenhaimer, A.~Li, K.~Das, and V.~Chirayath,
  ``\BIBforeignlanguage{en}{Cloud detection algorithm for multi-modal satellite
  imagery using convolutional neural-networks ({CNN})},''
  \emph{\BIBforeignlanguage{en}{Remote Sensing of Environment}}, vol. 237, p.
  111446, Feb. 2020.

\bibitem{shendryk_deep_2019}
Y.~Shendryk, Y.~Rist, C.~Ticehurst, and P.~Thorburn, ``Deep learning for
  multi-modal classification of cloud, shadow and land cover scenes in
  {PlanetScope} and {Sentinel}-2 imagery,'' \emph{ISPRS Journal of
  Photogrammetry and Remote Sensing}, vol. 157, pp. 124--136, Nov. 2019.

\bibitem{tote_evaluation_2018}
\BIBentryALTinterwordspacing
C.~Toté, E.~Swinnen, S.~Sterckx, S.~Adriaensen, I.~Benhadj, M.-D. Iordache,
  L.~Bertels, G.~Kirches, K.~Stelzer, W.~Dierckx, L.~Van~den Heuvel,
  D.~Clarijs, and F.~Niro, ``Evaluation of {PROBA}-{V} {Collection} 1:
  {Refined} {Radiometry}, {Geometry}, and {Cloud} {Screening},'' \emph{Remote
  Sensing}, vol.~10, no.~9, p. 1375, Aug. 2018. [Online]. Available:
  \url{https://www.mdpi.com/2072-4292/10/9/1375}
\BIBentrySTDinterwordspacing

\bibitem{QRSLiangBook}
S.~Liang, \emph{Quantitative Remote Sensing of Land Surfaces}.\hskip 1em plus
  0.5em minus 0.4em\relax USA: Wiley-Interscience, 2003.

\bibitem{mandanici_preliminary_2016}
E.~Mandanici and G.~Bitelli, ``\BIBforeignlanguage{en}{Preliminary {Comparison}
  of {Sentinel}-2 and {Landsat} 8 {Imagery} for a {Combined} {Use}},''
  \emph{\BIBforeignlanguage{en}{Remote Sensing}}, vol.~8, no.~12, p. 1014, Dec.
  2016.

\bibitem{revel_sentinel-2a_2019}
C.~Revel, V.~Lonjou, S.~Marcq, C.~Desjardins, B.~Fougnie, C.~C.-D. Luche,
  N.~Guilleminot, A.-S. Lacamp, E.~Lourme, C.~Miquel, and X.~Lenot,
  ``Sentinel-{2A} and {2B} absolute calibration monitoring,'' \emph{European
  Journal of Remote Sensing}, vol.~52, no.~1, pp. 122--137, Jan. 2019.

\bibitem{Zhao2018JSTARS}
Y.~{Zhao}, L.~{Ma}, C.~{Li}, C.~{Gao}, N.~{Wang}, and L.~{Tang}, ``Radiometric
  cross-calibration of landsat-8/oli and gf-1/pms sensors using an instrumented
  sand site,'' \emph{IEEE Journal of Selected Topics in Applied Earth
  Observations and Remote Sensing}, vol.~11, no.~10, pp. 3822--3829, Oct 2018.

\bibitem{houborg_cubesat_2018}
\BIBentryALTinterwordspacing
R.~Houborg and M.~F. McCabe, ``\BIBforeignlanguage{en}{A {Cubesat} enabled
  {Spatio}-{Temporal} {Enhancement} {Method} ({CESTEM}) utilizing {Planet},
  {Landsat} and {MODIS} data},'' \emph{\BIBforeignlanguage{en}{Remote Sensing
  of Environment}}, vol. 209, pp. 211--226, May 2018. [Online]. Available:
  \url{https://linkinghub.elsevier.com/retrieve/pii/S0034425718300786}
\BIBentrySTDinterwordspacing

\bibitem{claverie_harmonized_2018}
M.~Claverie, J.~Ju, J.~G. Masek, J.~L. Dungan, E.~F. Vermote, J.-C. Roger,
  S.~V. Skakun, and C.~Justice, ``The {Harmonized} {Landsat} and {Sentinel}-2
  surface reflectance data set,'' \emph{Remote Sensing of Environment}, vol.
  219, pp. 145--161, Dec. 2018.

\bibitem{zhang_characterization_2018}
\BIBentryALTinterwordspacing
H.~K. Zhang, D.~P. Roy, L.~Yan, Z.~Li, H.~Huang, E.~Vermote, S.~Skakun, and
  J.-C. Roger, ``\BIBforeignlanguage{en}{Characterization of {Sentinel}-{2A}
  and {Landsat}-8 top of atmosphere, surface, and nadir {BRDF} adjusted
  reflectance and {NDVI} differences},'' \emph{\BIBforeignlanguage{en}{Remote
  Sensing of Environment}}, vol. 215, pp. 482--494, Sep. 2018. [Online].
  Available:
  \url{http://www.sciencedirect.com/science/article/pii/S0034425718301883}
\BIBentrySTDinterwordspacing

\bibitem{scheffler_spectral_2020}
D.~Scheffler, D.~Frantz, and K.~Segl, ``Spectral harmonization and red edge
  prediction of {Landsat}-8 to {Sentinel}-2 using land cover optimized
  multivariate regressors,'' \emph{Remote Sensing of Environment}, vol. 241, p.
  111723, May 2020.

\bibitem{digitalimproc}
R.~C. Gonzalez and R.~E. Woods, \emph{Digital Image Processing (3rd
  Edition)}.\hskip 1em plus 0.5em minus 0.4em\relax USA: Prentice-Hall, Inc.,
  2006.

\bibitem{inamdar_multidimensional_2008}
S.~Inamdar, F.~Bovolo, L.~Bruzzone, and S.~Chaudhuri, ``Multidimensional
  {Probability} {Density} {Function} {Matching} for {Preprocessing} of
  {Multitemporal} {Remote} {Sensing} {Images},'' \emph{IEEE Transactions on
  Geoscience and Remote Sensing}, vol.~46, no.~4, pp. 1243--1252, Apr. 2008.

\bibitem{tuia_graph_2013}
D.~Tuia, J.~Mu{\~n}oz-Mar{\'i}, L.~G{\'o}mez-Chova, and J.~Malo, ``Graph
  {Matching} for {Adaptation} in {Remote} {Sensing},'' \emph{IEEE Transactions
  on Geoscience and Remote Sensing}, vol.~51, no.~1, pp. 329--341, Jan 2013.

\bibitem{tuia_semisupervised_2014}
D.~Tuia, M.~Volpi, M.~Trolliet, and G.~Camps-Valls, ``Semisupervised {Manifold}
  {Alignment} of {Multimodal} {Remote} {Sensing} {Images},'' \emph{IEEE
  Transactions on Geoscience and Remote Sensing}, vol.~52, no.~12, pp.
  7708--7720, Dec. 2014.

\bibitem{goodfellow_2014}
I.~Goodfellow, J.~Pouget-Abadie, M.~Mirza, B.~Xu, D.~Warde-Farley, S.~Ozair,
  A.~Courville, and Y.~Bengio, ``Generative adversarial nets,'' in
  \emph{Advances in Neural Information Processing Systems 27}.\hskip 1em plus
  0.5em minus 0.4em\relax Curran Associates, Inc., 2014, pp. 2672--2680.

\bibitem{ganin_domain-adversarial_2016}
\BIBentryALTinterwordspacing
Y.~Ganin, E.~Ustinova, H.~Ajakan, P.~Germain, H.~Larochelle, F.~Laviolette,
  M.~March, and V.~Lempitsky, ``Domain-{Adversarial} {Training} of {Neural}
  {Networks},'' \emph{Journal of Machine Learning Research}, vol.~17, no.~59,
  pp. 1--35, 2016. [Online]. Available:
  \url{http://jmlr.org/papers/v17/15-239.html}
\BIBentrySTDinterwordspacing

\bibitem{tzeng_adversarial_2017}
E.~Tzeng, J.~Hoffman, K.~Saenko, and T.~Darrell, ``Adversarial {Discriminative}
  {Domain} {Adaptation},'' in \emph{{CVPR} 2017}, Jul. 2017, pp. 2962--2971.

\bibitem{liu_unsupervised_2017}
\BIBentryALTinterwordspacing
M.-Y. Liu, T.~Breuel, and J.~Kautz, ``Unsupervised {Image}-to-{Image}
  {Translation} {Networks},'' in \emph{Advances in {Neural} {Information}
  {Processing} {Systems} 30}, I.~Guyon, U.~V. Luxburg, S.~Bengio, H.~Wallach,
  R.~Fergus, S.~Vishwanathan, and R.~Garnett, Eds.\hskip 1em plus 0.5em minus
  0.4em\relax Curran Associates, Inc., 2017, pp. 700--708. [Online]. Available:
  \url{http://papers.nips.cc/paper/6672-unsupervised-image-to-image-translation-networks.pdf}
\BIBentrySTDinterwordspacing

\bibitem{bousmalis_unsupervised_2017}
K.~Bousmalis, N.~Silberman, D.~Dohan, D.~Erhan, and D.~Krishnan, ``Unsupervised
  {Pixel}-{Level} {Domain} {Adaptation} with {Generative} {Adversarial}
  {Networks},'' in \emph{{CVPR} 2017}, Jul. 2017, pp. 95--104.

\bibitem{elshamli_domain_2017}
A.~Elshamli, G.~W. Taylor, A.~Berg, and S.~Areibi, ``Domain {Adaptation}
  {Using} {Representation} {Learning} for the {Classification} of {Remote}
  {Sensing} {Images},'' \emph{IEEE Journal of Selected Topics in Applied Earth
  Observations and Remote Sensing}, vol.~10, no.~9, pp. 4198--4209, Sep. 2017.

\bibitem{song_domain_2019}
S.~Song, H.~Yu, Z.~Miao, Q.~Zhang, Y.~Lin, and S.~Wang, ``Domain {Adaptation}
  for {Convolutional} {Neural} {Networks}-{Based} {Remote} {Sensing} {Scene}
  {Classification},'' \emph{IEEE Geoscience and Remote Sensing Letters},
  vol.~16, no.~8, pp. 1324--1328, Aug. 2019.

\bibitem{tasar_colormapgan_2020}
O.~Tasar, S.~L. Happy, Y.~Tarabalka, and P.~Alliez, ``{ColorMapGAN}:
  {Unsupervised} {Domain} {Adaptation} for {Semantic} {Segmentation} {Using}
  {Color} {Mapping} {Generative} {Adversarial} {Networks},'' \emph{IEEE
  Transactions on Geoscience and Remote Sensing}, pp. 1--16, 2020.

\bibitem{koga_method_2020}
Y.~Koga, H.~Miyazaki, and R.~Shibasaki, ``\BIBforeignlanguage{en}{A {Method}
  for {Vehicle} {Detection} in {High}-{Resolution} {Satellite} {Images} that
  {Uses} a {Region}-{Based} {Object} {Detector} and {Unsupervised} {Domain}
  {Adaptation}},'' \emph{\BIBforeignlanguage{en}{Remote Sensing}}, vol.~12,
  no.~3, p. 575, Jan. 2020.

\bibitem{benjdira_unsupervised_2019}
B.~Benjdira, Y.~Bazi, A.~Koubaa, and K.~Ouni,
  ``\BIBforeignlanguage{en}{Unsupervised {Domain} {Adaptation} {Using}
  {Generative} {Adversarial} {Networks} for {Semantic} {Segmentation} of
  {Aerial} {Images}},'' \emph{\BIBforeignlanguage{en}{Remote Sensing}},
  vol.~11, no.~11, p. 1369, Jan. 2019.

\bibitem{ye_sar_2019}
F.~Ye, W.~Luo, M.~Dong, H.~He, and W.~Min, ``{SAR} {Image} {Retrieval} {Based}
  on {Unsupervised} {Domain} {Adaptation} and {Clustering},'' \emph{IEEE
  Geoscience and Remote Sensing Letters}, vol.~16, no.~9, pp. 1482--1486, Sep.
  2019.

\bibitem{Wang19_J-GRSL_8825802}
L.~{Wang}, X.~{Xu}, Y.~{Yu}, R.~{Yang}, R.~{Gui}, Z.~{Xu}, and F.~{Pu},
  ``Sar-to-optical image translation using supervised cycle-consistent
  adversarial networks,'' \emph{IEEE Access}, vol.~7, pp. 129\,136--129\,149,
  2019.

\bibitem{Liu18_C-IGARSS_8517866}
L.~{Liu}, Z.~{Pan}, X.~{Qiu}, and L.~{Peng}, ``Sar target classification with
  cyclegan transferred simulated samples,'' in \emph{IGARSS 2018 - 2018 IEEE
  International Geoscience and Remote Sensing Symposium}, July 2018, pp.
  4411--4414.

\bibitem{pan_survey_2010}
S.~J. Pan and Q.~Yang, ``A {Survey} on {Transfer} {Learning},'' \emph{IEEE
  Transactions on Knowledge and Data Engineering}, vol.~22, no.~10, pp.
  1345--1359, Oct. 2010.

\bibitem{isola2017image}
P.~Isola, J.-Y. Zhu, T.~Zhou, and A.~A. Efros, ``Image-to-image translation
  with conditional adversarial networks,'' in \emph{{CVPR}, 2017}, 2017.

\bibitem{Mescheder2018ICML}
L.~Mescheder, S.~Nowozin, and A.~Geiger, ``Which training methods for gans do
  actually converge?'' in \emph{International Conference on Machine Learning
  (ICML)}, 2018.

\bibitem{Denaro20JSTARS}
L.~{Denaro} and C.~{Lin}, ``Hybrid canonical correlation analysis and
  regression for radiometric normalization of cross-sensor satellite imagery,''
  \emph{IEEE Journal of Selected Topics in Applied Earth Observations and
  Remote Sensing}, vol.~13, pp. 976--986, 2020.

\bibitem{geological_survey_l8_2016-1}
\BIBentryALTinterwordspacing
U.~Geological~Survey, ``L8 {Biome} {Cloud} {Validation} {Masks},'' \emph{U.S.
  Geological Survey}, no. data release, 2016. [Online]. Available:
  \url{http://doi.org/10.5066/F7251GDH}
\BIBentrySTDinterwordspacing

\bibitem{foga_cloud_2017}
S.~Foga, P.~L. Scaramuzza, S.~Guo, Z.~Zhu, R.~D. Dilley, T.~Beckmann, G.~L.
  Schmidt, J.~L. Dwyer, M.~J. Hughes, and B.~Laue, ``Cloud detection algorithm
  comparison and validation for operational {Landsat} data products,''
  \emph{Remote Sensing of Environment}, vol. 194, no. Suppl. C, pp. 379--390,
  Jun. 2017.

\bibitem{hughes_automated_2014}
M.~J. Hughes and D.~J. Hayes, ``\BIBforeignlanguage{en}{Automated {Detection}
  of {Cloud} and {Cloud} {Shadow} in {Single}-{Date} {Landsat} {Imagery}
  {Using} {Neural} {Networks} and {Spatial} {Post}-{Processing}},''
  \emph{\BIBforeignlanguage{en}{Remote Sensing}}, vol.~6, no.~6, pp.
  4907--4926, May 2014.

\bibitem{gomez-chova_cloud_2017}
L.~Gómez-Chova, G.~Mateo-García, J.~Muñoz-Marí, and G.~Camps-Valls, ``Cloud
  detection machine learning algorithms for {PROBA}-{V},'' in \emph{{IGARSS}
  2017}, Jul. 2017, pp. 2251--2254.

\bibitem{Wolters18}
\BIBentryALTinterwordspacing
E.~Wolters, W.~Dierckx, M.-D. Iordache, and E.~Swinnen, ``{PROBA-V products
  user manual},'' QWG, Tech. Rep. Technical Note, 16/03/2018, March 2018.
  [Online]. Available:
  \url{http://www.vito-eodata.be/PDF/image/PROBAV-Products_User_Manual.pdf}
\BIBentrySTDinterwordspacing

\bibitem{ToteIGARSS18}
C.~{Toté} and E.~{Swinnen}, ``{Extending the Spot/Vegetation--Proba-V Archive
  with Sentinel-3: a Preliminary Evaluation},'' in \emph{{IGARSS 2018 - 2018
  IEEE International Geoscience and Remote Sensing Symposium}}, 2018, pp.
  8707--8710.

\bibitem{sterckx_orbit_2016}
\BIBentryALTinterwordspacing
S.~Sterckx, S.~Adriaensen, W.~Dierckx, and M.~Bouvet,
  ``\BIBforeignlanguage{en}{In-{Orbit} {Radiometric} {Calibration} and
  {Stability} {Monitoring} of the {PROBA}-{V} {Instrument}},''
  \emph{\BIBforeignlanguage{en}{Remote Sensing}}, vol.~8, no.~7, p. 546, Jul.
  2016. [Online]. Available: \url{https://www.mdpi.com/2072-4292/8/7/546}
\BIBentrySTDinterwordspacing

\bibitem{KingmaB14}
D.~P. Kingma and J.~Ba, ``Adam: {A} method for stochastic optimization,'' in
  \emph{3rd International Conference on Learning Representations, {ICLR} 2015,
  San Diego, CA, USA, May 7-9, 2015, Conference Track Proceedings}, 2015, pp.
  1--13.

\bibitem{chollet_xception_2017}
F.~Chollet, ``Xception: {Deep} {Learning} with {Depthwise} {Separable}
  {Convolutions},'' in \emph{2017 {IEEE} {Conference} on {Computer} {Vision}
  and {Pattern} {Recognition} ({CVPR})}, Jul. 2017, pp. 1800--1807.

\end{thebibliography}

\end{document}